\documentclass{aa}

\usepackage[T1]{fontenc}
\usepackage[utf8]{inputenc}
\usepackage{graphicx}
\usepackage{xcolor}
\usepackage{array}
\usepackage{needspace}
\usepackage{txfonts}
\usepackage[hidelinks]{hyperref}

\AtBeginDocument{%
  \nolinenumbers
  \let\linenumbers\nolinenumbers
}

\begin{document}

\title{What determines the properties of Population III binaries? }
\titlerunning{Properties of Population III binaries}

\author{Neha Singh\inst{1}\corrauth{neha.singh@uib.es}
        \and Tomasz Bulik\inst{2}}
\authorrunning{N. Singh \& T. Bulik}

\institute{IAC3, Universitat de les Illes Balears, Cra. de Valldemossa km 7.5, 07122 Palma, Spain
           \email{neha.singh@uib.es}
           \and
           Astronomical Observatory, University of Warsaw,
           Al. Ujazdowskie 4, 00478 Warszawa, Poland
           \email{tb@astrouw.edu.pl}}

\date{}

\abstract{Population~III binaries may provide a gravitational-wave probe of
the first stellar populations. Uncertain initial binary parameter
distributions and evolutionary prescriptions complicate predictions of
their remnant masses, birth orbits and formation spins. Varying these model evolutionary parameters (MEPs) jointly is needed to
identify which MEPs shape each property and whether their effects
persist among binaries that merge. We use \textsc{StarTrack} to evolve
$2\times10^6$ initial binaries per model across 1152 configurations
spanning eight MEPs. To our knowledge, this is the largest grid of
Population~III isolated-binary population-synthesis models to date by
number of distinct parameter configurations. We compare the properties
and correlations of all binary black holes (BBHs) with those of systems
that merge within the adopted age of the Universe. The all-BBH (merging-BBH) model medians span
$34.3$--$171.0\,M_{\odot}$ ($23.1$--$94.8\,M_{\odot}$) in total mass,
$0.039$--$0.106$ ($0.018$--$0.129$) in effective aligned spin
$\chi_{\mathrm{eff}}$ at formation, and $0.075$--$0.639$
($0.039$--$0.983$) in birth eccentricity. 
Merger selection also changes correlations between BBH properties and
the combinations of MEP values and prescriptions associated with them.
We use variance-based sensitivity analysis to quantify how the initial binary parameter
distributions and evolutionary prescriptions affect the predicted BBH
properties. For the all-BBH population, initial binary parameter
distributions have the largest influence on mass-ratio, chirp-mass and birth-orbit
statistics, whereas natal-kick magnitude leads for delay times and spins.
For merging BBHs, initial distributions remain central to mass variation,
with substantial contributions from common-envelope efficiency to mass
ratios, orbits and delays; the angular-momentum-loss coefficient leads for median $\chi_{\mathrm{eff}}$, whereas natal-kick magnitude leads for median $\chi_{\mathrm{p}}$.
These results identify population-dependent responses to evolutionary
assumptions within the tested grid. They describe intrinsic populations
without cosmic star-formation weighting or detector selection effects, which we defer to a future paper.}

\keywords{binaries: general -- gravitational waves -- methods: numerical --
          methods: statistical -- stars: black holes -- stars: Population III}

\maketitle

\section{Introduction}
\label{sec:introduction}
Population~III (Pop~III) stars are the first generation of stars, formed
from primordial gas. Without cooling by metals and dust, their birth clouds
follow a different thermal evolution from present-day star-forming gas.
Simulations generally predict a mass distribution weighted towards higher
masses than in present-day stellar populations \citep{KlessenGlover2023}.
Their radiation and nucleosynthetic products initiated the ionisation and
chemical enrichment of the intergalactic medium, modifying the conditions
for subsequent star formation \citep{BrommLarson2004}. 
The initial properties of Pop~III stellar populations, including their initial
mass function, binary fraction and orbital-parameter distributions, remain uncertain. These uncertainties
propagate into predictions for their compact remnants. Binary black holes (BBHs) 
formed from Pop~III stars may merge long after their progenitors formed, providing
a possible gravitational-wave probe of primordial stellar populations. Relating
their masses and spins to the progenitor population requires models of binary 
evolution and the associated delay-time distribution \citep{KlessenGlover2023}.

\citet{BelczynskiEtAl2017} investigated Pop~III compact-binary formation with
\textsc{StarTrack} \citep{BelczynskiEtAl2002,BelczynskiEtAl2008},
combining initial binary properties derived from
$N$-body calculations based on \citet{RyuEtAl2016} with stellar-evolution
prescriptions adapted to metal-free stars.  In their comparison with earlier predictions,
\citet{BelczynskiEtAl2017} demonstrated that the predicted BBH merger rates
depend strongly on the initial binary parameter distributions and
the treatment of binary interactions. 

Previous studies have explored the dependence of Pop~III binary populations
on uncertain model evolutionary parameters (MEPs). \citet{TanikawaEtAl2022} varied the initial
mass function, initial binary properties, stellar-evolution prescriptions
and tidal evolution in models spanning Pop~I, II and III stars.
\citet{HijikawaEtAl2022} investigated common-envelope evolution and the
initial mass function through 288 parameter combinations for massive
Pop~III binaries. \citet{CostaEtAl2023} examined how the initial mass
function and initial binary parameter distributions affect BBH properties
and formation pathways, while \citet{SantoliquidoEtAl2023} quantified
the dependence of Pop~III BBH merger rates on these distributions and the
adopted star-formation history. 

Parameter studies have also investigated whether models of stellar and binary
evolution can reproduce the properties and merger rates inferred from
LIGO--Virgo--KAGRA
\citep{AasiEtAl2015LIGO,AcerneseEtAl2015Virgo,AkutsuEtAl2021KAGRA} observations
\citep[e.g.][]{AbbottEtAl2019GWTC1Population,AbbottEtAl2024GWTC21,AbbottEtAl2023GWTC3Population}.
\citet{TanikawaEtAl2021} varied stellar
winds, natal kicks and the lower limits of initial mass ratios and
separations, comparing predicted BBH properties and merger rates with
observations. Their pair-instability mass-gap channel produced
GW190521-like masses \citep{AbbottEtAl2020GW190521} but underpredicted the inferred event rate.
\citet{Tanikawa2024} examined how convective overshooting and binary
interactions affect envelope retention and the formation of massive
remnants, comparing combined Pop~I/II/III populations with the mass and
merger-rate distributions inferred from GWTC-3
\citep{AbbottEtAl2023GWTC3,AbbottEtAl2023GWTC3Population}.
\citet{TanikawaEtAl2026} obtained GW231123-like masses and merger rates
\citep{AbacEtAl2025GW231123}
with inefficient convective overshooting and a
$^{12}\mathrm{C}(\alpha,\gamma)^{16}\mathrm{O}$ reaction rate at least
$2\sigma$ below their adopted standard, assuming Pop~I/II-like initial
orbital distributions. These constraints depend on the assumed formation
channel and do not establish a Pop~III origin for the observed events.
The mapping between the progenitor population and its compact remnants also
depends on uncertain prescriptions for mass transfer, common-envelope
evolution and compact-object formation \citep{BelczynskiEtAl2017}.
Varying these assumptions jointly is therefore necessary to quantify their
contributions to the predicted population properties.

The main goal of this work is to quantify how the adopted initial binary
parameter distributions and evolutionary prescriptions affect the predicted
properties of Pop~III compact binaries. We use variance-based sensitivity
analysis to identify the MEPs that most strongly influence each property
and quantify their contributions to variation across models.
Our simulations use the \textsc{StarTrack} framework of
\citet{BelczynskiEtAl2017}, updated to include black-hole spin evolution. Although the
initial binary parameter distributions are informed by $N$-body calculations, we
evolve the systems as isolated field binaries, without further
dynamical encounters with other stars. We generate a
complete grid of 1152 models resulting from the combinations of values and prescriptions for eight MEPs, evolving
$2\times10^6$ initial binaries per model, for a total of
$2.304\times10^9$ simulated binary systems.
To our knowledge, this is the largest grid of Pop~III isolated-binary
population-synthesis models to date by number of distinct parameter
configurations. We focus on BBH masses, orbital properties, delay times and
spins, comparing the all-BBH population with the subset merging within the
adopted age of the Universe. We also examine within-model correlations
between binary properties and how these change after merger selection.

The paper is organised as follows. Section~\ref{sec:generation} describes
how we generate the model grid. Section~\ref{sec:results} defines the compact-binary properties and presents
the population comparisons and correlations.
Section~\ref{sec:variance-analysis} motivates the variance-based sensitivity
analysis, describes its method and examines the contributions of the varied
MEPs. We discuss the interpretation and limitations in
Sect.~\ref{sec:discussion} and conclude in Sect.~\ref{sec:conclusions}.
The mathematical derivations are given in Appendix~\ref{app:variance-analysis}.

\section{Population synthesis and catalogue construction}
\label{sec:generation}

\begin{table*}[t]
\caption{Model evolutionary parameters varied in the Pop~III binary-population grid.}
\label{tab:input-parameters}
\centering
\small
\setlength{\tabcolsep}{5pt}
\renewcommand{\arraystretch}{1.15}
\begin{tabular}{@{}>{\raggedright\arraybackslash}p{0.24\textwidth}
                  >{\raggedright\arraybackslash}p{0.14\textwidth}
                  >{\raggedright\arraybackslash}p{0.57\textwidth}@{}}
\hline\hline
MEP & Values or prescription codes & Physical meaning or prescription \\
\hline
Common-envelope efficiency, $\alpha$ & $0.5,\ 1,\ 1.5$ &
Efficiency with which released orbital energy ejects the envelope in the
$\alpha$--$\lambda$ energy prescription. \\[3pt]
Angular-momentum loss, $\beta$ & $0.5,\ 1,\ 2$ &
Coefficient setting the orbital angular momentum removed by non-accreted
matter in the mass-transfer branches that use this parameter. \\[3pt]
Magnetic braking (\texttt{MB}) & $0,\ 3$ &
0: disabled; 3: the \citet{IvanovaTaam2003} prescription for stars satisfying
the magnetic-braking criteria. \\[3pt]
Remnant prescription (\texttt{REMNANT}) & $2,\ 3$ &
2: rapid supernova prescription \citep{FryerEtAl2012}; 3: the same prescription
with pulsational pair-instability mass loss and pair-instability disruption
\citep{BelczynskiEtAl2016PI}. \\[3pt]
Initial binary parameter distributions (\texttt{SS}) & $4,\ 5$ &
4: P1 prescription, with relatively close initial binaries; 5: P4 prescription,
with a broader separation distribution extending to much wider binaries.
Both use initial mass, mass-ratio, separation and eccentricity distributions
based on the $N$-body calculations of \citet{RyuEtAl2016}, following their
use in binary population synthesis by \citet{BelczynskiEtAl2017}. \\[3pt]
Natal-kick magnitude (\texttt{kick}) & $6,\ 8,\ 9,\ 10$ &
6: Maxwellian kicks reduced by fallback;
8: Maxwellian kicks for neutron stars, scaled by $2.5\,M_{\odot}/M_{\rm BH}$
for black holes;
9: a kick relation based on the ejecta-to-remnant mass ratio
\citep{BrayEldridge2016};
10: the vector sum of a fallback-reduced mass-ejection component and an
unreduced neutrino component. \\[3pt]
Fallback threshold (\texttt{kickback}) & $0,\ 1$ &
Full kicks are retained below a fallback mass of $0.2\,M_{\odot}$ (0) or
$0.6\,M_{\odot}$ (1). This setting affects the fallback-dependent components
of kick prescriptions 6 and 10. \\[3pt]
Natal-kick direction (\texttt{polar}) & $0,\ 1$ &
0: isotropic; 1: aligned with the positive pre-supernova orbital
angular-momentum direction for a bound binary. \\
\hline
\end{tabular}
\tablefoot{All MEPs are dimensionless coefficients or prescription codes.
P1 and P4 are the labels used in the implementation for the initial-population
prescriptions. The Maxwellian draws in kick prescriptions 6 and 8 use a
one-dimensional dispersion of $265\,\mathrm{km\,s^{-1}}$
\citep{HobbsEtAl2005}. The fallback threshold controls whether reduction
is applied; the reduction itself depends on the fallback fraction.}
\end{table*}

\subsection{Description of StarTrack}
\label{sec:startrack}
We generate the binary populations with \textsc{StarTrack}, a rapid
population-synthesis code that follows single and binary stellar evolution
using parametrised stellar models and prescriptions for binary interactions
\citep{BelczynskiEtAl2002,BelczynskiEtAl2008}. The code treats mass transfer,
common-envelope evolution, tides, stellar winds, magnetic braking,
compact-remnant formation and natal kicks, and orbital decay through
gravitational-wave emission. Our Pop~III calculations follow the framework
of \citet{BelczynskiEtAl2017}, suppressing stellar winds and limiting radial
expansion to approximately match that of $Z=0$ stellar models. We use initial
binary properties derived from the dynamical calculations of \citet{RyuEtAl2016}.
\citet{BelczynskiEtAl2017} summarise these initial binary parameter
distributions in their Table~2 and illustrate them in their Figs.~6--9.

Ryu et al. followed the dynamics
of small groups of Pop~III stars, initially on nearly Keplerian orbits in a
disc, including mutual gravitational interactions, gas dynamical friction
and the gravitational potential of a uniform gas background. The two
configurations adopted by \citet{BelczynskiEtAl2017} contained five
stars each, with initial spatial scales of approximately $2000\,\mathrm{au}$
and $10$--$20\,\mathrm{au}$. The resulting distributions of primary mass,
mass ratio, orbital separation and eccentricity supplied the initial binary
properties for population synthesis. These initial-binary
prescriptions are also used in this work.

Our implementation additionally includes black-hole spin evolution.
Each black hole is assigned an initial dimensionless spin of $0.05$ at
formation; its spin can subsequently evolve through accretion.

\subsection{Model evolutionary parameters}
\label{sec:input-parameters}
We vary eight MEPs describing the initial binary parameter distributions and the
treatment of binary evolution and compact-object formation
(Table~\ref{tab:input-parameters}). Each MEP is specified by a numerical value or a choice of prescription. The initial-population prescription (\texttt{SS}) sets the distributions of
stellar masses, mass ratios, orbital separations and eccentricities at the
start of binary evolution. The common-envelope efficiency $\alpha$
parametrises how efficiently orbital energy released during inspiral ejects
the shared stellar envelope. The coefficient $\beta$ sets the orbital
angular momentum carried away by matter lost from the binary during
non-conservative mass transfer onto a non-degenerate accretor. Magnetic
braking (\texttt{MB}) describes the loss of stellar rotational angular
momentum through winds coupled to the stellar magnetic field. As the gas
moves outward, magnetic stresses maintain approximate corotation with the
star beyond its surface. This increases the effective lever arm of the
outflow, allowing the escaping gas to carry away more angular momentum
and slow the stellar rotation \citep{IvanovaTaam2003}. In \textsc{StarTrack}, this process is
represented by the parametrised angular-momentum-loss prescription of
\citet{IvanovaTaam2003}, which depends on stellar radius and rotation rate
without explicitly evolving the stellar magnetic field.

The remnant prescription (\texttt{REMNANT}) determines the type and mass of
the compact object left by stellar collapse, including whether pair
instability causes additional mass loss or complete disruption
\citep{FryerEtAl2012,BelczynskiEtAl2016PI}. The natal-kick prescription
(\texttt{kick}) sets the magnitude of the velocity impulse imparted to the
newly formed remnant. The fallback threshold (\texttt{kickback}) specifies
the minimum mass of material falling back onto the remnant for kick
reduction to apply in fallback-dependent kick components. The kick-direction
prescription (\texttt{polar}) specifies the orientation of this impulse
relative to the pre-collapse orbital angular momentum.

We include all combinations of the tested MEP values or prescriptions listed in
Table~\ref{tab:input-parameters}, giving $3\times3\times2\times2\times2\times4\times2\times2
=1152$ distinct model configurations. For each configuration, we evolve
$2\times10^6$ initial binaries, corresponding to $2.304\times10^9$ binary
realisations across the grid. The numerical labels distinguish discrete
prescriptions without implying a physical ordering.

\begin{table*}[t]
\caption{Compact-binary counts in the catalogue across the model grid.}
\label{tab:binary-counts}
\centering
\begin{tabular}{llrrr}
\hline\hline
Binary type & Sample & \multicolumn{2}{c}{Binary counts per model} & Nonempty / total models \\
\cline{3-4}
 & & Minimum & Maximum & \\
\hline
BBH & \emph{All}     & 2550 & 1\,305\,176 & 1152/1152 \\
BBH & \emph{Merging} & 0    & 249\,344    & 1053/1152 \\
BNS & \emph{All}     & 0    & 479         & 685/1152  \\
BNS & \emph{Merging} & 0    & 8           & 193/1152  \\
NSBH & \emph{All}     & 11 & 49\,505     & 1152/1152 \\
NSBH & \emph{Merging} & 0  & 386         & 609/1152  \\
\hline
\end{tabular}
\tablefoot{Counts refer to compact binaries included in the catalogue. All 1152 models enter each
row, including models with zero binaries in the stated sample. The minimum
and maximum describe counts per model; nonempty models contain at
least one binary in that sample. \emph{All} imposes no additional delay cut;
\emph{merging} is the subset satisfying the total-delay criterion defined in
Sect.~\ref{sec:catalogue-selection}.}
\end{table*}

For computational convenience, the 1152 configurations were generated in
two sets of 576 models. Model labels in the figures retain the set
identifier and the model number within that set (\texttt{set:model}).
The statistical analyses treat the two sets jointly as a single catalogue,
while retaining the individual model configurations.

\subsection{Catalogue assembly and selection}
\label{sec:catalogue-selection}

Table~\ref{tab:binary-counts} summarises the numbers of simulated binary
black holes (BBHs), binary neutron stars (BNSs) and neutron-star--black-hole
binaries (NSBHs) across the grid. We classify compact
remnants with masses above $2.5\,M_{\odot}$ as black holes
and those with masses $\leq2.5\,M_{\odot}$ as neutron stars.
The \emph{all} and \emph{merging} samples use this remnant classification
and the total-delay criterion below.
The total delay is $t_{\mathrm{delay}}=t_{\mathrm{evol}}+t_{\mathrm{mer}}$, where
$t_{\mathrm{evol}}$ is the time from zero-age main sequence (ZAMS) binary formation
until both compact remnants have formed, and $t_{\mathrm{mer}}$ is the
subsequent time to merger. All three times are expressed in Myr.

The \emph{all-BBH} population comprises all BBHs in a given model,
without an additional delay cut. The \emph{merging-BBH} population is
the subset with total formation-to-merger delay $t_{\mathrm{delay}}\leq
t_{\mathrm H}$, where $t_{\mathrm{delay}}$ is measured from the formation
of the initial stellar binary and $t_{\mathrm H}=13\,786.9\,\mathrm{Myr}$ 
is the adopted Hubble-time threshold.
This selection specifies a duration after binary formation, 
rather than a merger epoch in cosmic history.

\section{Compact-binary populations}
\label{sec:results}

\subsection{Definitions of compact-binary properties}
\label{sec:properties}

We describe the masses, orbital properties and spins of the compact
binaries in the catalogue, with the detailed comparisons focused
on binary black holes (BBHs).

\subsubsection{Masses}
\label{sec:properties-masses}
For each compact binary, we order the component masses as $m_1\geq m_2$,
where the heavier remnant need not descend from the initially more massive
star. The total mass $M_{\mathrm{tot}}$, mass ratio $q$ and chirp mass $\mathcal{M}$ are
\begin{equation}
 M_{\mathrm{tot}}=m_1+m_2,\qquad q=\frac{m_2}{m_1},\qquad
 \mathcal{M}=\frac{(m_1m_2)^{3/5}}{(m_1+m_2)^{1/5}}.
 \label{eq:binary-masses}
\end{equation}
Here $0<q\leq1$.
All masses are source-frame values, expressed in solar masses
($M_{\odot}$).

\subsubsection{Orbital properties}
\label{sec:properties-orbits}
The simulation records the semi-major axis $a_{\mathrm{birth}}$ and eccentricity $e_{\mathrm{birth}}$ at
the compact-binary formation snapshot, after both remnants have formed.
We refer to this eccentricity as the birth eccentricity.

For the BBH correlation analysis, we also estimate the eccentricity
$e_{0.1}$ when twice the mean orbital frequency $f_{\mathrm{orb}}$
reaches the source-frame reference value $2f_{\mathrm{orb}}=0.1\,\mathrm{Hz}$.
For eccentric binaries, this convention does not identify the harmonic carrying the greatest
gravitational-wave power. Starting from the saved formation values of
$a_{\mathrm{birth}}$ and $e_{\mathrm{birth}}$, we use the orbit-averaged quadrupole evolution of two point
masses derived by \citet{Peters1964}, retaining its full eccentricity
dependence. This calculation evolves the orbit without spin feedback;
it does not evolve the spin quantities defined below. We estimate
$e_{0.1}$ for the \emph{all-BBH} population in the 1152 models.

\subsubsection{Spins}
\label{sec:properties-spins}
For each black hole $i=1,2$, the dimensionless spin magnitude is $\chi_i=cS_i/(Gm_i^2)$, 
where $S_i$ is the magnitude of its spin
angular momentum, $c$ is the speed of light and $G$ is the gravitational
constant. If $\theta_i$ is the
angle between its spin and the orbital angular momentum at BBH formation,
the effective aligned spin $\chi_{\mathrm{eff}}$ \citep{GerosaEtAl2021} is
\begin{equation}
 \chi_{\mathrm{eff}}=
 \frac{m_1\chi_1\cos\theta_1+m_2\chi_2\cos\theta_2}{m_1+m_2}.
 \label{eq:effective-spin}
\end{equation}
The spin labels follow the same mass ordering as $m_1$ and $m_2$.
Positive and negative values indicate a net
mass-weighted spin projection along and opposite to the orbital angular
momentum, respectively. The effective precession parameter $\chi_{\mathrm{p}}$ \citep{SchmidtEtAl2015} is
\begin{equation}
 \chi_{\mathrm{p}}=\max\!\left[
 \chi_1\sin\theta_1,\,
 \frac{q(4q+3)}{4+3q}\chi_2\sin\theta_2\right],
 \label{eq:precession-spin}
\end{equation}
where $\max$ selects the larger of the two weighted contributions
\citep[see also Eq.~14 of][]{GerosaEtAl2021}. The mass-ratio factor
multiplies the lighter black hole's contribution.
The effective aligned spin $\chi_{\mathrm{eff}}$ is approximately
conserved during isolated inspiral, whereas $\chi_{\mathrm{p}}$ can vary as the
spin directions precess \citep{GerosaEtAl2021}. We report only the
values of $\chi_{\mathrm{eff}}$ and $\chi_{\mathrm{p}}$ at BBH birth, calculated
from the stored spin magnitudes and tilts, for both the \emph{all-BBH} and
\emph{merging-BBH} populations.

\begin{figure*}[!tp]
  \centering
  \includegraphics[width=\textwidth]{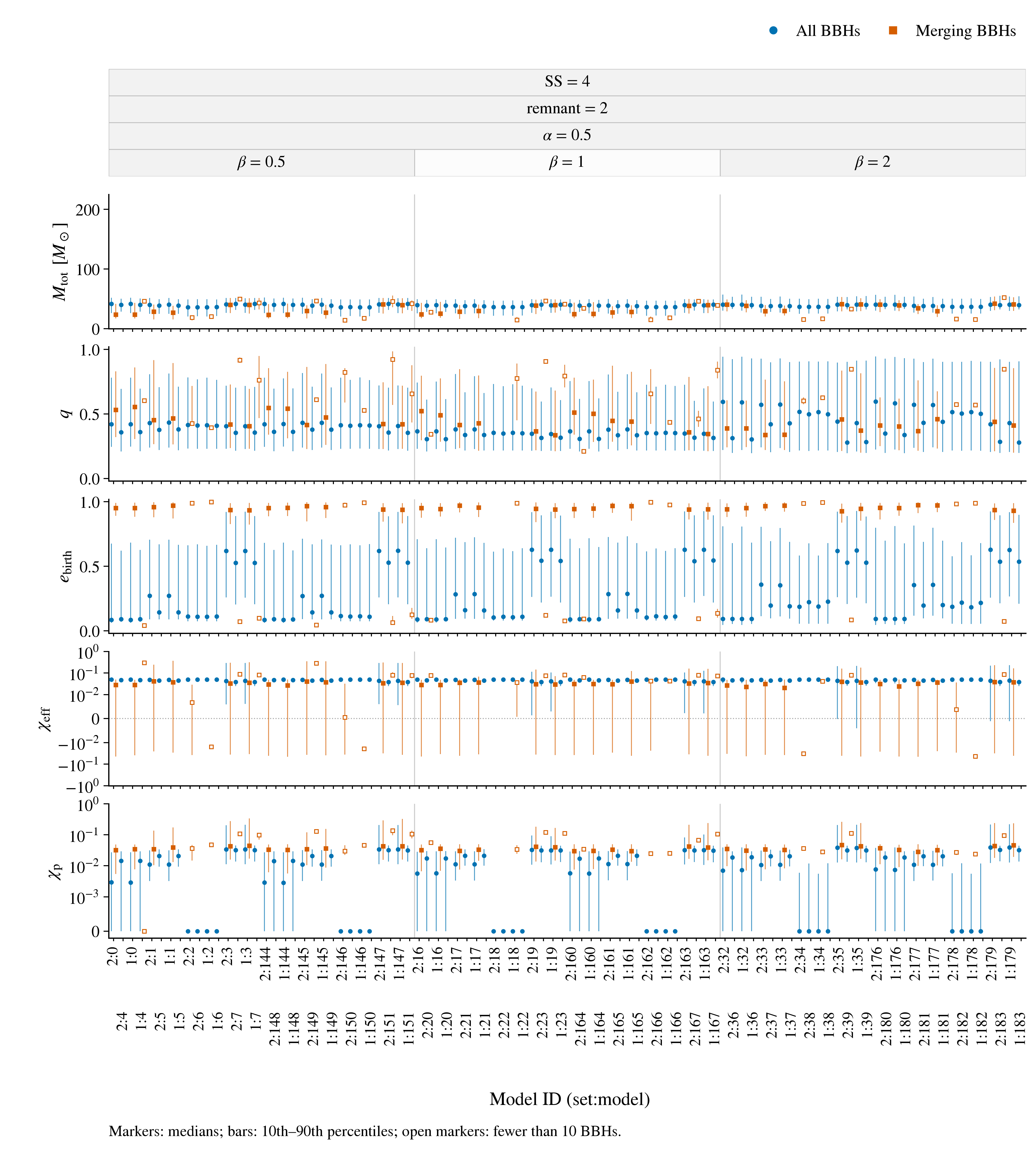}
  \caption{Model-wise BBH summaries for SS=4, remnant=2 and $\alpha=0.5$
    (section 1). Blue circles show the \emph{all-BBH} population; orange squares
    show the \emph{merging-BBH} population. From top to bottom, the panels
    show $M_{\mathrm{tot}}$, $q$, $e_{\mathrm{birth}}$, $\chi_{\mathrm{eff}}$
    and $\chi_{\mathrm{p}}$. Total masses are expressed in solar masses
    ($M_{\odot}$).
    Markers give within-model medians and bars span the 10th--90th percentiles
    of the binary population. Open markers indicate fewer than ten BBHs.
    Plots covering the full model grid and the underlying data will be
    made available on Zenodo.}
  \label{fig:bbh-section-01}
\end{figure*}

\begin{figure*}[!tp]
  \centering
  \includegraphics[width=\textwidth]{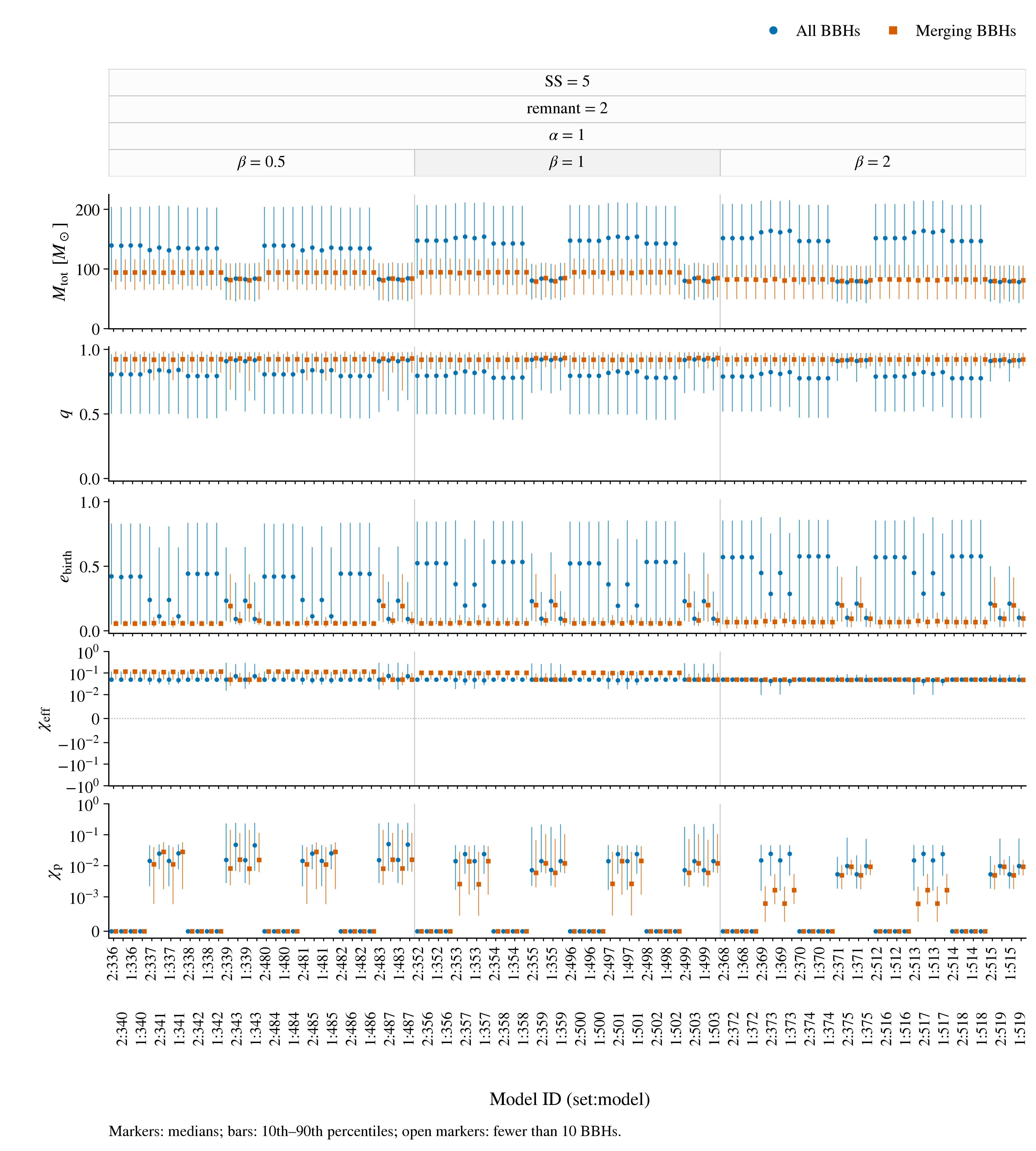}
  \caption{Model-wise BBH summaries for SS=5, remnant=2 and $\alpha=1$
    (section 8). Populations, markers and percentile ranges follow
    Fig.~\ref{fig:bbh-section-01}. Plots covering the full model grid and
    the underlying data will be made available on Zenodo.}
  \label{fig:bbh-section-08}
\end{figure*}

\subsection{Population properties across the model grid}
\label{sec:results-properties}

Across the explored grid, isolated Pop~III evolution produces BBH
populations with a broad range of total masses $M_{\mathrm{tot}}$, mass ratios $q$
and birth eccentricities $e_{\mathrm{birth}}$. Requiring merger within $t_{\mathrm H}$ changes these populations:
in most models, the merging binaries have lower median $M_{\mathrm{tot}}$ and
higher median $q$ than the \emph{all-BBH} population.
Their $e_{\mathrm{birth}}$ values show a less uniform pattern, with some model
groups producing merging populations with high median $e_{\mathrm{birth}}$.

We present the results for the full model grid in terms of within-model
medians and percentile ranges. The \emph{all-BBH} summaries include all
1152 models; unless stated otherwise, merger summaries use the 992 models
with at least ten merging BBHs. Figures~\ref{fig:bbh-section-01}
and~\ref{fig:bbh-section-08} provide examples from the model grid.

Before imposing the total-delay cut, all 1152 models contain BBHs and
NSBHs, while 685 contain BNSs (Table~\ref{tab:binary-counts}). Requiring
$t_{\mathrm{delay}}\leq t_{\mathrm H}$ leaves nonempty \emph{merging}
samples in 1053 models for BBHs, 609 for NSBHs and 193 for BNSs.
Thus, 99 models contain BBHs but no merging BBHs, and 543 contain NSBHs
but no merging NSBHs within this time limit. Of the 1053 nonempty
\emph{merging-BBH} samples, 61 contain fewer than ten binaries.

\textbf{Total mass:} Across the grid, \emph{all-BBH} median $M_{\mathrm{tot}}$ spans
$34.3$--$171.0\,M_{\odot}$, approximately a factor of five.
The \emph{merging-BBH} medians of $M_{\mathrm{tot}}$ span $23.1$--$94.8\,M_{\odot}$
and are lower than their \emph{all-BBH} counterparts in 792 of
the 992 paired comparisons. Including the low-count merger
samples extends the median range down to $14.0\,M_{\odot}$.

The largest model-level 90th percentiles of $M_{\mathrm{tot}}$ across the grid are
approximately $217\,M_{\odot}$ for \emph{all-BBH} and
$118\,M_{\odot}$ for \emph{merging-BBH}. The extended high-mass
intervals of \emph{all-BBH} are visible in
Fig.~\ref{fig:bbh-section-08}.
    
The range of median $M_{\mathrm{tot}}$ across models narrows after
merger selection, particularly at its high-mass end. For example, the
upper panels show \emph{all-BBH} medians near
$36$--$42\,M_{\odot}$ in Fig.~\ref{fig:bbh-section-01}, compared
with $78$--$164\,M_{\odot}$ in Fig.~\ref{fig:bbh-section-08}.
In the latter example, merger medians occupy the narrower range
$79$--$95\,M_{\odot}$, illustrating the reduction in median $M_{\mathrm{tot}}$
found in many models across the grid.

\textbf{Mass ratio:} Across the grid, model medians of $q$ span $0.278$--$0.962$ for \emph{all-BBH}
and $0.326$--$0.973$ for \emph{merging-BBH}.
Merger selection raises the median $q$ in 940 of the
992 paired comparisons.

The grid contains both nearly equal-mass and substantially
unequal-mass merger populations. For example, in Fig.~\ref{fig:bbh-section-01},
\emph{all-BBH} medians of $q$ span $0.28$--$0.60$, and the 36 merger samples with at least ten
binaries have medians $0.34$--$0.55$, while in Fig.~\ref{fig:bbh-section-08}, the corresponding ranges are
$0.78$--$0.92$ and $0.92$--$0.93$. In the latter group, merger medians cluster
near equal masses, although their percentile bars extend to
lower $q$.

\textbf{Birth eccentricity:} Across the grid, median $e_{\mathrm{birth}}$ spans $0.075$--$0.639$ for
\emph{all-BBH} and $0.039$--$0.983$ for \emph{merging-BBH}.
Although the merger median is lower in 864 of the 992 paired
comparisons, some model groups select very eccentric systems.
 
The Hubble-time selection does not uniformly favour low
$e_{\mathrm{birth}}$: merger populations with both low and high
median $e_{\mathrm{birth}}$ occur across the grid. For example,
in the third panel of Fig.~\ref{fig:bbh-section-01}, the 36 merger samples with at
least ten binaries have median $e_{\mathrm{birth}}$ in
$0.93$--$0.97$, compared with $0.08$--$0.63$ for
\emph{all-BBH} across that group. In
Fig.~\ref{fig:bbh-section-08}, merger medians instead lie at
$0.057$--$0.201$, compared with $0.09$--$0.58$ for
\emph{all-BBH}. These two groups illustrate the contrasting
eccentricity regimes present in the full grid.

\textbf{Formation spins:} Across the grid, median $\chi_{\mathrm{eff}}$ spans $0.039$--$0.106$ for
\emph{all-BBH} and $0.018$--$0.129$ for \emph{merging-BBH}.
These medians are positive, although some within-model
percentile intervals extend to negative values.

Median $\chi_{\mathrm{p}}$ ranges from zero to $0.119$ for
\emph{all-BBH} and from zero to $0.094$ for \emph{merging-BBH},
with larger values in the within-model tails. For example, the
bottom panels of Figs.~\ref{fig:bbh-section-01}
and~\ref{fig:bbh-section-08} show groups of zero medians alongside nonzero
medians of order $10^{-2}$, with some percentile bars extending
above $0.1$.
These spin results are conditional on the adopted
spin evolution and fixed initial BH spin of 0.05.

Across the full grid, all seven negative merger medians of $\chi_{\mathrm{eff}}$
belong to samples of only one to three binaries; these do not
establish a well-populated negative-spin subpopulation.
Including all low-count merger samples expands the merger-median
$\chi_{\mathrm{eff}}$ interval to $-0.044$--$0.305$. Examples of these negative medians
appear as open orange markers in the fourth panel of
Fig.~\ref{fig:bbh-section-01}. In both displayed groups, the
samples with at least ten binaries illustrate the positive
medians found throughout the grid under this count restriction.

Figures~\ref{fig:bbh-section-01} and~\ref{fig:bbh-section-08} were
selected to show contrasting examples of the variation across the full
model grid: mergers with lower $M_{\mathrm{tot}}$, lower $q$ and high
$e_{\mathrm{birth}}$ in the first group, and higher $M_{\mathrm{tot}}$, $q$ closer
to unity and lower $e_{\mathrm{birth}}$ in the second.
These figures also show how merger selection changes the population
within a model. Both SS and $\alpha$ differ between these groups, so
their contrast alone does not isolate either MEP's effect.

\begin{figure*}[!tp]
	\centering
  \includegraphics[width=0.7\textwidth]{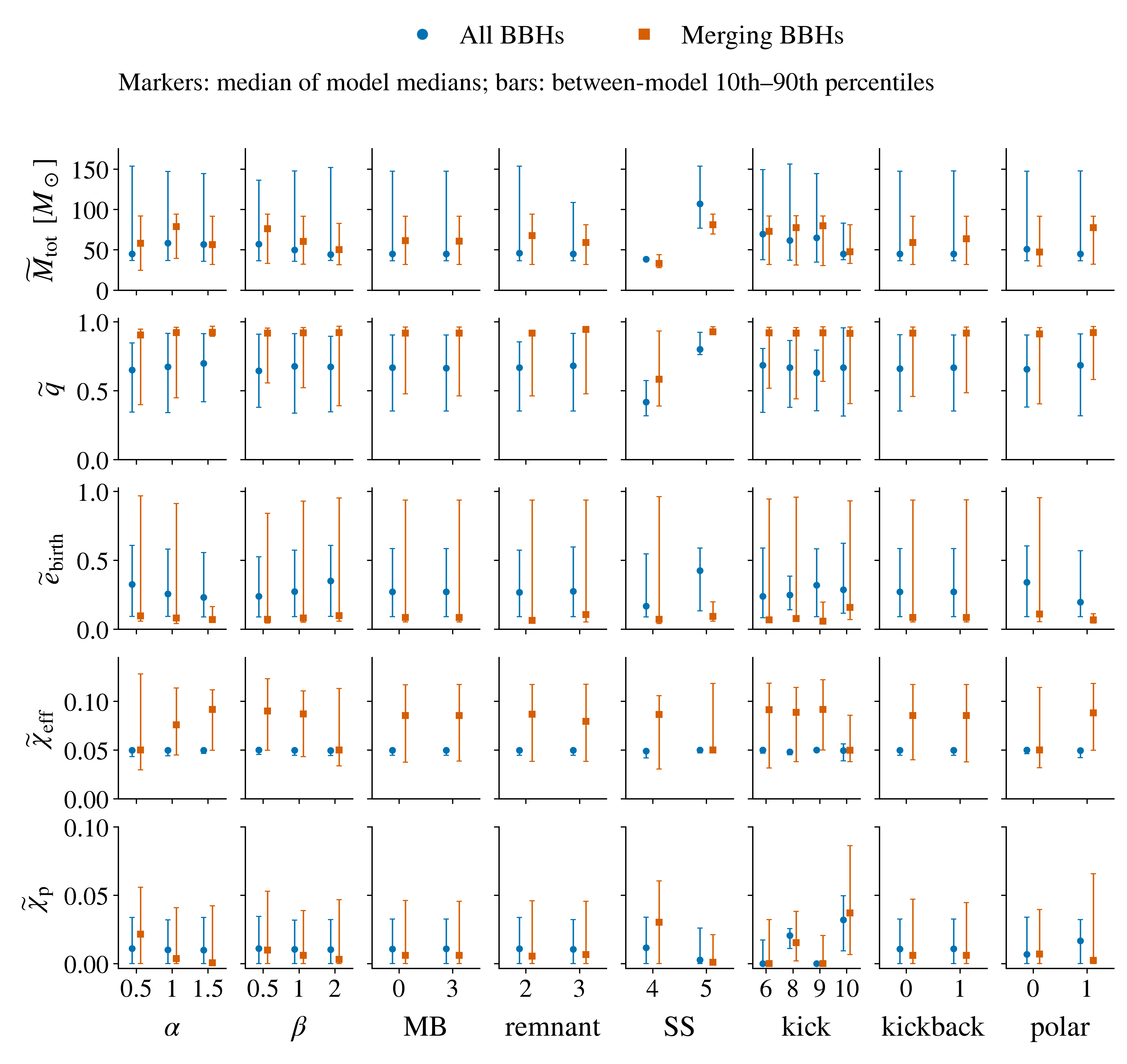}
  \caption{Dependence of population summaries on MEP values or prescriptions for the \emph{all-BBH} population
    (blue circles, shifted left) and the \emph{merging-BBH} population
    (orange squares, shifted right). Horizontal axes show numerical values or prescription codes. Paired markers refer to the same value or prescription. Total masses are expressed in solar masses ($M_{\odot}$).
    Tildes denote within-model medians. Markers give medians of these
    within-model medians; bars span their between-model 10th--90th percentiles,
    with equal model weights, not confidence intervals. All 1152 models
    contribute to the \emph{all-BBH} summaries; 1053 nonempty models contribute
    after the total-delay cut $\leq t_{\mathrm H}$. The populations overlap.
    After merger selection, the combinations of other MEP values and prescriptions can differ between the plotted values or prescriptions,
    so the \emph{merging-BBH} summaries are descriptive and conditional.}
  \label{fig:bbh-factor-responses}
\end{figure*}

Figure~\ref{fig:bbh-factor-responses} extends these comparisons by grouping
models according to each of the eight varied MEPs. It summarises the
within-model medians of $M_{\mathrm{tot}}$, $q$, $e_{\mathrm{birth}}$,
$\chi_{\mathrm{eff}}$ and $\chi_{\mathrm{p}}$. At each MEP value or prescription, the marker
is the median of these model medians, and the bar spans their 10th--90th
percentiles across the other MEP values or prescriptions, with equal weight per model.
The bars describe between-model variation, not within-model binary ranges
or confidence intervals. This figure includes all 1152 \emph{all-BBH}
models and all 1053 nonempty \emph{merging-BBH} models, including the 61
with fewer than ten mergers. The grouped medians of $M_{\mathrm{tot}}$ and $q$ are higher
for SS=5 than for SS=4 in both populations. The percentile bars show the spread in model medians among models
that share the same value or prescription for one MEP but differ in their other MEP values or prescriptions. Such comparisons reveal shifts across MEP values or prescriptions but do not quantify each MEP's
contribution to the total variation. They can also conceal interactions,
in which the change associated with one MEP depends on the values or prescriptions of the other MEPs.
These limitations motivate the variance-based analysis in
Sect.~\ref{sec:variance-analysis}, which quantifies contributions involving
each MEP while accounting for interactions and the selected model ensemble.

\subsection{Correlations between binary properties}
\label{sec:results-correlations}
\begin{figure*}[!tp]
  \centering
  \includegraphics[width=0.8\textwidth]{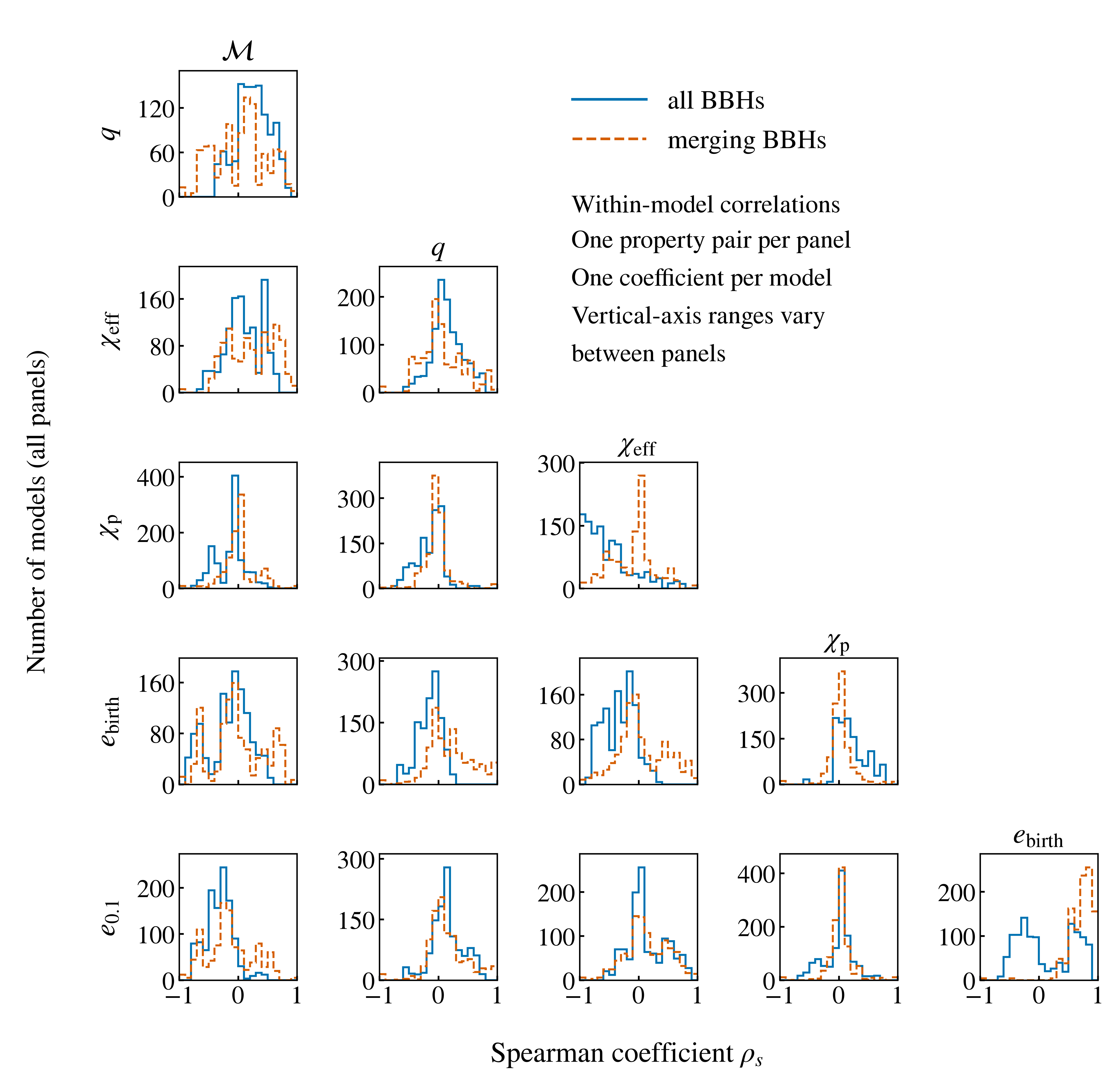}
  \caption{Distributions of within-model Spearman correlations for the six
    selected BBH properties: $\mathcal{M}$, $q$, $\chi_{\mathrm{eff}}$,
    $\chi_{\mathrm{p}}$, $e_{\mathrm{birth}}$ and $e_{0.1}$. Blue solid and orange dashed outlines show the
    \emph{all-BBH} and \emph{merging-BBH} populations, respectively, with one
    coefficient per valid model.
    Top and left symbols identify the paired properties; every horizontal axis
    shows the Spearman coefficient $\rho_s$. Each panel shows a coefficient
    histogram, not a joint response density. Vertical-axis ranges vary between panels. $\chi_{\mathrm{eff}}$ and $\chi_{\mathrm{p}}$ are formation
    values; $e_{0.1}$ is the Peters estimate at source-frame
    $2f_{\mathrm{orb}}=0.1\,\mathrm{Hz}$.}
  \label{fig:bbh-correlation-corner}
\end{figure*}

Figure~\ref{fig:bbh-correlation-corner} shows how associations between binary
properties vary across the model grid and with merger selection. For each
property pair, we calculate the Spearman rank coefficient $\rho_s$ separately
within each model and population, assigning average ranks to tied values.
Positive and negative coefficients indicate increasing and decreasing
monotonic associations, respectively. Each histogram contains one coefficient
per valid model, with equal model weights; it does not show correlations
between model medians or a population formed by pooling binaries across models.
The numerical summaries below are medians of these within-model coefficients.

Every pair has 1152 valid \emph{all-BBH} coefficients and 1020 valid
\emph{merging-BBH} coefficients. The latter exclude 99 empty and 33
single-binary samples but include 28 models with two to nine mergers, unlike
the count-qualified comparisons in Sect.~\ref{sec:results-properties}. 

The spin pair provides a clear example of the effect of merger selection.
For \emph{all-BBH}, the $\chi_{\mathrm{eff}}$--$\chi_{\mathrm{p}}$
distribution is predominantly negative, with median $\rho_s=-0.636$.
In the \emph{merging-BBH} sample it is concentrated much closer to zero,
with median $-0.005$, although both positive and negative correlations remain.
Thus the negative association between the two formation-spin measures in
the full population is much less prominent among systems selected to merge
within $t_{\mathrm H}$. The correlations of $\chi_{\mathrm{p}}$ with
$\mathcal{M}$, $q$ and $e_{0.1}$ are concentrated near zero in both
populations. In these correlations, $e_{0.1}$ is obtained by evolving the
orbit to $2f_{\mathrm{orb}}=0.1\,\mathrm{Hz}$ using the Peters equations,
while $\chi_{\mathrm{eff}}$ and $\chi_{\mathrm{p}}$ are evaluated at BBH
formation.

The $e_{\mathrm{birth}}$--$e_{0.1}$ pair shows the opposite change: its
median coefficient is $-0.020$ for \emph{all-BBH} but $0.774$ for
\emph{merging-BBH}. The full-population distribution spans both signs,
whereas the merger-selected distribution is concentrated at positive
coefficients. Within the merger-selected models, binaries with larger
birth eccentricities therefore tend to retain larger eccentricities at
the reference frequency. This is a statement about rank ordering, not
about the magnitude of the residual eccentricity or its detectability.

The mass-related pairs show that merger selection does not shift all
correlations in the same direction. The median $\mathcal{M}$--$q$
coefficient decreases from $0.239$ to $0.104$, while that of
$\mathcal{M}$--$\chi_{\mathrm{eff}}$ increases from $0.062$ to $0.223$.
The median $\mathcal{M}$--$e_{0.1}$ coefficient remains negative,
changing from $-0.300$ to $-0.188$, whereas the median
$q$--$e_{\mathrm{birth}}$ coefficient changes sign, from $-0.115$ to
$0.179$. In each case the values refer to \emph{all-BBH} and
\emph{merging-BBH}, respectively. The broad distributions show that these
median trends are not shared by every model. The changes in these
coefficient distributions show that the merger cut affects relationships
between properties as well as the individual-property distributions described above.

The $q$--$\chi_{\mathrm{eff}}$ pair illustrates how a near-zero ensemble
median can conceal differences that could help distinguish models. Its
median coefficient is $0.004$ across the 1020 valid \emph{merging-BBH}
models. Restricting this comparison to the 992 models with at least ten
mergers, 480 have $|\rho_s|<0.2$, while 136 have $\rho_s<-0.3$ and 251
have $\rho_s>0.3$. These descriptive thresholds do not establish the
absence or statistical significance of an association. The negative-tail
models contain 271--133589 mergers each, so this tail persists beyond
the sparsest samples. All 136 have angular-momentum-loss coefficient
$\beta=1$ or $2$, with none at $\beta=0.5$ in this selected grid; however,
both $\beta=1$ and $2$ also occur among models with positive correlations.

We next examine the MEP values or prescriptions associated with the
positive and negative tails of the distributions of within-model
correlation coefficients across models, using the thresholds specified
below. We compare the \emph{all-BBH} populations of all 1152 models
with the \emph{merging-BBH} populations of the 992 models containing
at least ten mergers. The coefficients are calculated separately for
the two populations within each model. To distinguish changes caused
by selecting binaries from changes in the model ensemble, we also
compare the two sets of coefficients for the same 992 configurations.
Prescription labels follow Table~\ref{tab:input-parameters}.

\textbf{Joint spin and mass-ratio correlations:}
For the \emph{all-BBH} population, the same 32 models satisfy both
correlation thresholds: $\rho_s(q,\chi_{\mathrm{eff}})<-0.3$ and
$\rho_s(\chi_{\mathrm{eff}},\chi_{\mathrm{p}})>0.3$.
All have SS=5 and kick=10. For the \emph{merging-BBH} population,
20 models meet both conditions; all have $\beta=1$, SS=5, REMNANT=2
and kick=8, and each contains more than $10^4$ merging BBHs.
All 32 original models have at least ten mergers, but none meets both
thresholds after merger selection; the 20 merger-tail models are
different configurations. The change in the associated kick prescription
therefore reflects a change in the correlations within models, rather
than the exclusion of configurations with too few mergers.
For the merger-tail models, $\beta=1$ assigns the non-accreted matter the binary's mean
specific orbital angular momentum in the mass-transfer prescription
for a non-degenerate accretor. The initial binary parameter distributions
SS=5 include a broad separation distribution extending to much wider
binaries than SS=4. REMNANT=2 uses the rapid supernova prescription
\citep{FryerEtAl2012}, without the pulsational pair-instability mass loss
and pair-instability disruption included in REMNANT=3. With kick=8,
black-hole natal-kick speeds are drawn from the adopted Maxwellian and
scaled by $2.5\,M_{\odot}/M_{\rm BH}$, without an additional fallback
reduction. Thus, for the same random draw, a more massive BH receives
a smaller kick. This mass-dependent recoil can couple remnant masses
to binary survival and changes in orbital orientation, which affect
the spin projections entering $\chi_{\mathrm{eff}}$ and
$\chi_{\mathrm{p}}$. It therefore provides a possible link between
the mass and spin correlations. In the \emph{all-BBH} tail, kick=10
instead combines recoil from asymmetric mass ejection, reduced by
fallback, with recoil from asymmetric neutrino emission, which is
unaffected by fallback in this prescription. Both prescriptions can produce recoil
even when fallback is complete, but with different mass dependence.

\textbf{Chirp mass and reference eccentricity:}
In the \emph{all-BBH} population, 28 models have
$\rho_s(\mathcal{M},e_{0.1})>0.3$; all have SS=5, REMNANT=3,
kick=10 and kicks along the pre-collapse orbital angular momentum
(polar=1). For merging BBHs, 183 models exceed the same threshold;
all have SS=5 and REMNANT=3 and contain at least 1000 mergers, but
they span all four kick-magnitude prescriptions and both kick-direction
prescriptions. All 28 original models contain at least ten mergers. When
restricted to merging BBHs, 11 remain above the correlation threshold
and 17 no longer exceed it; 172 additional models enter the positive tail. Thus, the association with the initial distributions and
pair-instability prescription persists, while the restriction to one
kick prescription and direction does not.
In the negative tail, $\rho_s(\mathcal{M},e_{0.1})<-0.5$ selects
226 \emph{all-BBH} models spanning both initial prescriptions and
all three common-envelope efficiencies. Of these, 178 have at least
ten mergers. On the same 992 configurations, the negative tail changes
from 178 models using all BBHs to 186 using merging BBHs, with only
36 models in both tails. All 186 merger-tail models have SS=4 and
$\alpha=1$ or $1.5$; 128 contain at least 1000 mergers.
In both populations, the positive tail combines initial binary parameter
distributions extending to wide separations (SS=5) with the rapid
supernova prescription including pulsational pair-instability mass loss
and pair-instability disruption (REMNANT=3). These additional processes
modify the remnant masses or prevent BH formation for affected progenitors
\citep{BelczynskiEtAl2016PI}. The merger-selected negative tail instead occurs for the
relatively close initial binaries of SS=4 and the two higher tested
common-envelope efficiencies. At fixed envelope binding energy, a larger
$\alpha$ requires less orbital energy to eject the envelope and hence
less orbital shrinkage in the energy prescription. It can therefore
change both the separations of surviving binaries and which systems
satisfy the merger-time condition. These values and prescriptions can alter the joint
distribution of remnant masses and birth orbits, from which subsequent
gravitational-wave evolution determines $e_{0.1}$.

\textbf{Chirp mass and precession spin:}
For the \emph{all-BBH} population, 694 models have
$|\rho_s(\mathcal{M},\chi_{\mathrm{p}})|<0.2$, while 44 have
$\rho_s>0.3$; all 44 have SS=5 and kick=8 or 10. For merging BBHs,
683 models satisfy the weak-correlation criterion and 156 have
$\rho_s>0.3$; all 156 have SS=5 and at least 1000 mergers.
All 44 original positive-tail models have at least ten mergers:
28 stay above the threshold, 16 fall below it and 128 additional
models enter the tail. The SS=5 association is therefore shared by
both populations, although the configurations contributing to the
tail change. Across the 992 models containing at least ten mergers,
the chirp-mass--precession-spin correlation coefficient is higher for
merging BBHs than for all BBHs in 818 models and lower in 174.
Holding the other seven MEPs fixed, changing SS=4 to SS=5
increases this coefficient in all 193 matched comparisons in which
both models contain at least 1000 mergers.
Physically, this change replaces the initial binary parameter
distributions of SS=4, with relatively close binaries, by those of SS=5,
which extend to much wider separations. The initial stellar masses,
mass ratios and eccentricities also change, so the matched comparisons
test the combined initial-distribution prescription. This comparison
between prescriptions is distinct from comparing all and merging BBHs
within one model. Changes in the initial distributions can
alter the mass-transfer histories and orbital configurations at stellar
collapse, and consequently the masses, spins and orbital tilts of the
systems that survive and merge. The precession parameter measures the
spin components in the orbital plane, with a mass-ratio weighting for
the lighter BH (Eq.~\ref{eq:precession-spin}); it has no direct dependence
on the absolute mass scale. The positive correlation therefore indicates
that larger chirp masses tend to accompany larger weighted in-plane
spin components within these models. This can reflect differences in
spin magnitudes, spin tilts and mass ratios, rather than a relation
between chirp mass and tilt alone. With the other seven MEPs held
fixed, changing from SS=4 to SS=5 consistently shifts the correlation
coefficient towards more positive values.

\textbf{Aligned spin and precession spin:}
For the \emph{all-BBH} population, 902 models have
$\rho_s(\chi_{\mathrm{eff}},\chi_{\mathrm{p}})<-0.3$, 132 have
$|\rho_s|<0.2$ and 64 have $\rho_s>0.3$.
All 64 positive-tail models have $\beta=0.5$
or $1$ and kick=10. In the \emph{merging-BBH} population, 291 models
lie below $-0.3$, 503 have $|\rho_s|<0.2$ and 116 lie above $0.3$.
All 116 positive-tail models have $\beta=0.5$ or $1$ together with
kick=8 or 10, and all contain at least 1000 mergers.
All 64 original positive-tail models have at least ten mergers:
32 remain above the threshold and 32 leave the tail; 84 additional
models enter it, all with kick=8. Among the 992 models containing at
least ten mergers, 742 have $\rho_s<-0.3$ using all BBHs and 291
using merging BBHs.
The coefficient increases after merger selection in 789 of these
992 models. Thus, the stronger representation of weak and positive
correlations is not solely due to excluding low-count merger models.
The two values of the angular-momentum-loss coefficient
remove less angular momentum than the remaining tested value, $\beta=2$,
for the same mass loss and binary configuration. They can therefore
change the orbit before stellar collapse. The two
kick prescriptions described above permit BH recoil without complete
suppression by fallback. The resulting recoil can change the orbital
plane relative to the spins and affect which binaries remain bound.
The resulting $\chi_{\mathrm{eff}}$ and $\chi_{\mathrm{p}}$ measure
spin components along the orbital angular momentum and in the orbital
plane, respectively. Their positive association means that binaries
with larger signed aligned-spin projections tend also to have larger
weighted in-plane spin components. Because both parameters depend on
spin magnitudes, orientations and mass ratio, this pattern reflects
their combined variation across the population.

\begin{figure*}[!tp]
  \centering
  \begin{minipage}[t]{0.49\textwidth}
    \vspace{0pt}
    \raggedright
    \textbf{(a)}\par\smallskip
    \includegraphics[width=\linewidth]{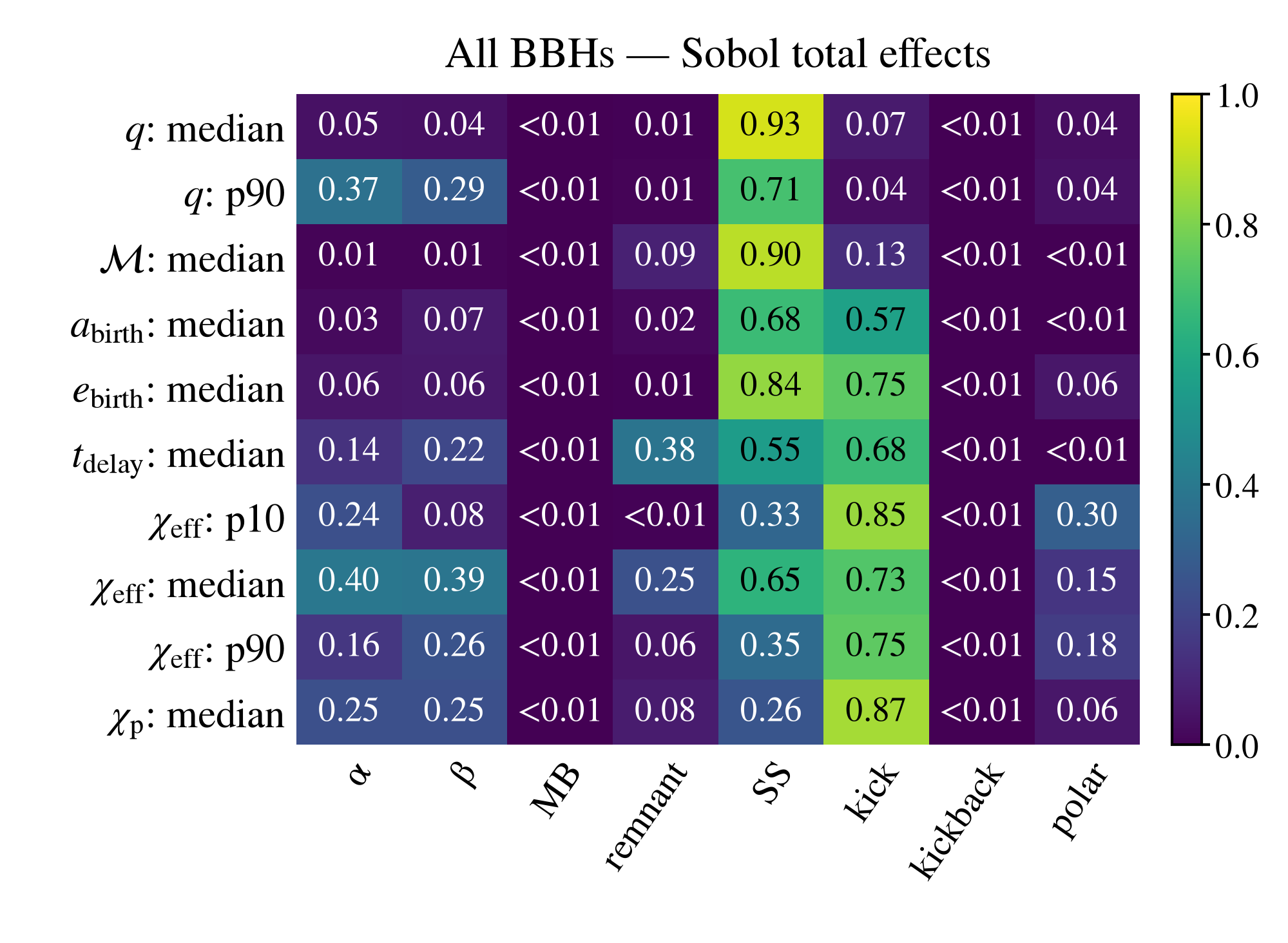}
  \end{minipage}\hfill
  \begin{minipage}[t]{0.49\textwidth}
    \vspace{0pt}
    \raggedright
    \textbf{(b)}\par\smallskip
    \includegraphics[width=\linewidth]{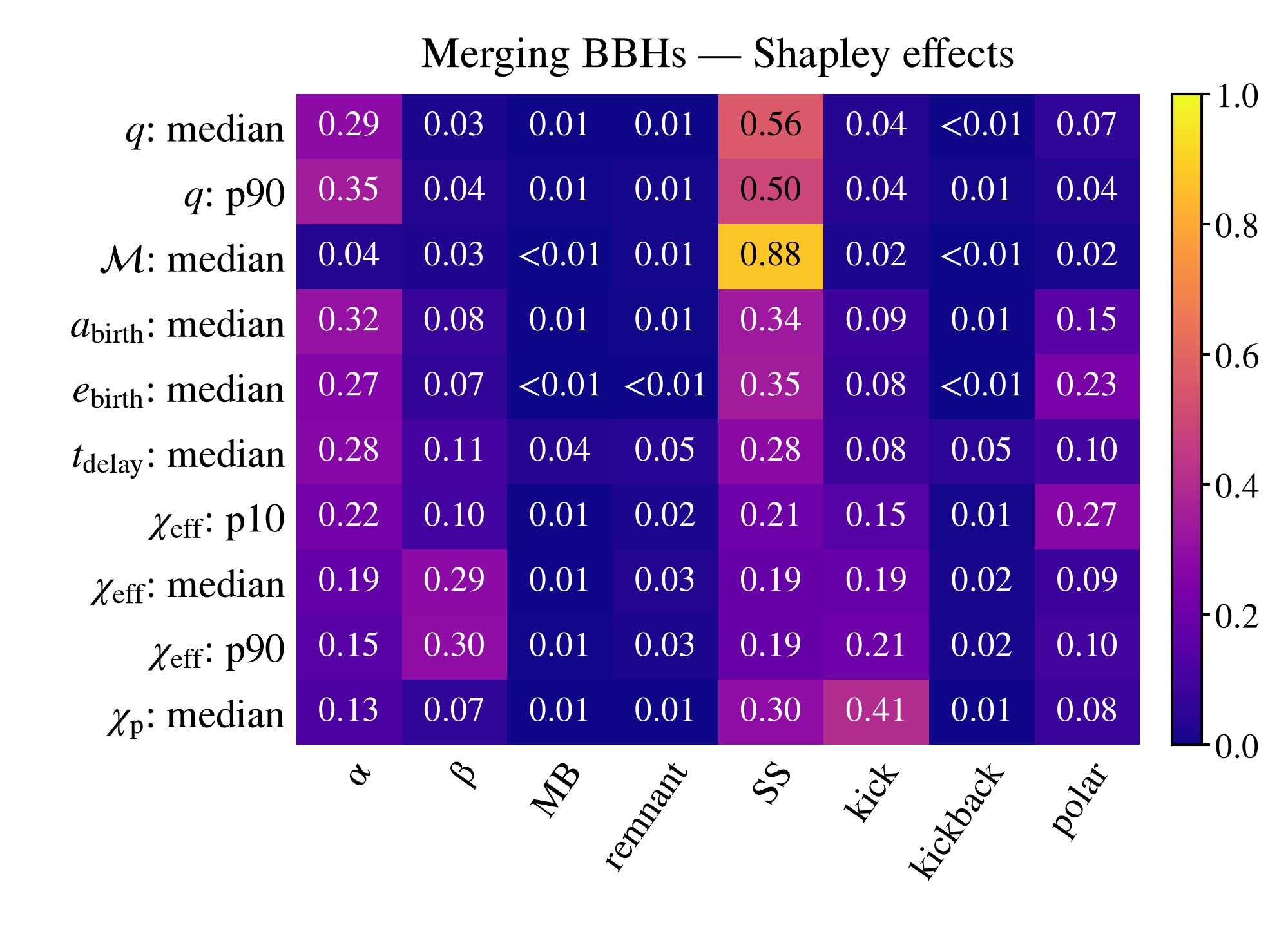}
  \end{minipage}
  \caption{Variance-based sensitivity of BBH response summaries to the varied
    MEPs. \textbf{(a)} Sobol total-effect indices for the indicated
    within-model summaries across the complete 1152-model \emph{all-BBH} grid.
    Each index includes the MEP's main effect and interactions; a row need
    not sum to one. \textbf{(b)} Variance-based Shapley effects for the indicated
    \emph{merging-BBH} summaries across the 1053 nonempty models; rows sum to
    one before rounding. In both panels, p10 and p90 denote the 10th and 90th
    percentiles. The panels use different sensitivity measures as well as
    different populations. Their response summaries are not identical to
    those shown in Fig.~\ref{fig:bbh-factor-responses}.}
  \label{fig:bbh-variance-heatmaps}
\end{figure*}

\textbf{Chirp mass and mass ratio:}
For the \emph{all-BBH} population, 44 of the 1152 models have
$\rho_s(\mathcal{M},q)<-0.3$; all have SS=5 and REMNANT=3.
Using the \emph{merging-BBH} population within each model, 228 of
the 992 models with at least ten mergers meet the same threshold;
all have SS=4, and 152 contain at least 1000 mergers.
All 44 original tail models have at least ten mergers, but none
remains below $-0.3$ after merger selection. Conversely, none of
the 228 merger-tail models meets this threshold in its all-BBH
population. The change from SS=5 to SS=4 in the tail therefore
persists when comparing the same model configurations; it is not
caused by excluding models with too few mergers.
The SS=4 prescription starts from relatively close binaries, whereas
SS=5 includes a much broader separation distribution; their initial
stellar-mass, mass-ratio and eccentricity distributions also differ.
These choices affect both the masses available to form BHs and the
mass-transfer histories that can change the mass ratio. A negative
$\mathcal{M}$--$q$ association means that larger chirp masses tend
to occur in binaries with more unequal component masses. 
The negative tail therefore reflects how total mass and mass ratio vary together
within the population. The merger-time condition selects binaries
according to their masses and birth orbits, so it can change this
joint distribution. The different initial prescriptions associated
with the two tails illustrate why the mass-ratio trend of the full
population need not carry over to its merging subset.

Observed population correlations could therefore help narrow the allowed
combinations of MEP values and prescriptions, provided that these differences persist after
weighting by the Pop~III star-formation history and merger delays and
accounting for detector selection effects, measurement uncertainties and other
formation channels. Detector selection effects arise because the
probability of detecting a BBH merger depends on its intrinsic properties,
distance and orientation, as well as the detectors' sensitivity. Consequently,
the detected sample can have different property distributions and correlations
from the intrinsic population. The strength of such constraints cannot be established
from the present coefficient distributions alone. For all the correlation
patterns discussed above, the shared MEP values or prescriptions identify associations
within the simulated populations but do not establish the mechanisms
responsible for their signs. Identifying those mechanisms requires examining
the binaries' evolutionary histories.
Models incorporating Pop~III star-formation histories and predictions for
the properties of detectable BBHs will be presented in a forthcoming paper
by Singh et al.\ (in preparation).

\section{Variance-based sensitivity analysis}
\label{sec:variance-analysis}

To quantify the MEP dependence illustrated in
Fig.~\ref{fig:bbh-factor-responses}, we analyse the variance of population
summaries across the model ensemble. The methods below account for both
interactions among MEPs and the model selection imposed by the merger cut.

A \emph{response} is a single numerical summary of a binary population
produced by one model configuration. For a specified property, summary
statistic and population selection, we denote this number by $y_m$, where
$m$ labels the model. For example, if the chosen response is the median
chirp mass $\mathcal{M}$ of the \emph{all-BBH} population, each model supplies
one value computed from its simulated BBHs. The corresponding
\emph{merging-BBH} response is computed separately from the subset satisfying
the total-delay cut. We analyse each response separately: its variance
measures the spread of these summary values across models, rather than
the spread of individual binary properties within a model. Each model
receives equal weight regardless of its binary count. These weights
describe the sampled grid rather than astrophysical occurrence probabilities.

For the complete 1152-model \emph{all-BBH} grid, we calculate Sobol total-effect
indices \citep{Sobol2001,Owen2013}. Each index includes the contribution
of an MEP and all interactions involving it.
For the \emph{merging-BBH} responses, excluding models with no qualifying
mergers removes configurations from the grid and can make the MEP values or prescriptions in the nonempty merger ensemble
statistically dependent. We therefore use variance-based
Shapley effects on this ensemble of models with nonempty merger samples \citep{SongEtAl2016,OwenPrieur2017}.
This analysis includes all 1053 nonempty merger models, including the
61 with fewer than ten mergers; the 992-model restriction used for most
population comparisons in Sect.~\ref{sec:results-properties} does not apply
here.
Appendix~\ref{app:variance-analysis} gives the mathematical definitions, 
derivations and treatment of interactions.

Figure~\ref{fig:bbh-variance-heatmaps} shows the sensitivity measures for
ten responses in each population: the within-model medians of
$q$, $\mathcal{M}$, $a_{\mathrm{birth}}$, $e_{\mathrm{birth}}$,
$t_{\mathrm{delay}}$, $\chi_{\mathrm{eff}}$ and $\chi_{\mathrm{p}}$;
the 90th percentile of $q$; and the 10th and 90th percentiles of
$\chi_{\mathrm{eff}}$. The medians describe typical population properties,
while the additional percentiles describe the upper part of the $q$
distribution and the lower and upper parts of the $\chi_{\mathrm{eff}}$
distribution. The descriptive comparisons in
Fig.~\ref{fig:bbh-factor-responses} use the separate set of five median
responses introduced in Sect.~\ref{sec:results-properties}.

The indices describe variation of model summaries over the specified
ensemble. They do not establish the direction of a response, statistical
significance or astrophysical causality. Their values depend on the tested
MEP values or prescriptions, model weights and nonlinear transformations of the response.

\subsection{Dependence on model evolutionary parameters}

\label{sec:results-inputs}
The relative importance of the MEPs depends on both the
population property and whether we consider its median or a percentile. In the
\emph{all-BBH} grid, the choice of initial binary parameter distributions
(\texttt{SS}) has the largest
Sobol total-effect index for the mass-ratio, chirp-mass and birth-orbit
responses, whereas the natal-kick magnitude prescription (\texttt{kick}) has the largest index for
the delay-time and spin responses
(Fig.~\ref{fig:bbh-variance-heatmaps}a). Thus, no single MEP dominates
all of the population summaries.

\textbf{Masses and mass ratios:}
For the \emph{all-BBH} population, the indices for the initial binary
parameter distributions are 0.93 for median
$q$ and 0.90 for median $\mathcal{M}$, substantially exceeding those
of the other MEPs. The upper mass-ratio percentile has a broader MEP
dependence: the initial distributions remain the leading MEP at 0.71,
but the total-effect indices of common-envelope efficiency ($\alpha$)
and angular-momentum loss during non-conservative mass transfer ($\beta$)
are 0.37 and 0.29, compared with
about 0.05 and 0.04 for median $q$. The median alone therefore does
not capture the sensitivity of the upper part of the mass-ratio
distribution. In the \emph{merging-BBH} ensemble
(Fig.~\ref{fig:bbh-variance-heatmaps}b), the choice of initial distributions receives 0.88 of the
Shapley allocation for median $\mathcal{M}$ and remains the leading
MEP for both mass-ratio summaries, with effects of 0.56 and 0.50
for the median and p90. Common-envelope efficiency is the next largest
contribution to these two responses, at 0.29 and 0.35. Within the
tested prescriptions, the initial binary parameter distributions therefore remain
central to the variation of mass summaries even when only mergers
are considered.

\textbf{Birth orbits and delay times:}
The \emph{all-BBH} median birth separation and eccentricity depend
strongly on both the initial binary parameter distributions and the
natal-kick magnitude prescription: their total-effect indices are 0.68 and
0.57 for $a_{\mathrm{birth}}$, and 0.84 and 0.75 for
$e_{\mathrm{birth}}$, respectively. These overlapping totals cannot
be treated as separate additive fractions; they indicate contributions
from interactions among MEPs, without identifying a particular
interacting pair. For median $t_{\mathrm{delay}}$, the kick magnitude
prescription has the largest index (0.68), followed by the initial
distributions (0.55) and the compact-remnant formation prescription
(\texttt{REMNANT}; 0.38). In the merger ensemble, the initial distributions
and common-envelope efficiency instead
receive the largest allocations for the three orbital and delay
summaries: 0.34 and 0.32 for median $a_{\mathrm{birth}}$, 0.35 and
0.27 for median $e_{\mathrm{birth}}$, and approximately 0.28 each
for median $t_{\mathrm{delay}}$. Natal-kick direction (\texttt{polar}) also contributes
to the birth-orbit summaries, with Shapley effects of 0.15 and 0.23
for separation and eccentricity. The predicted birth orbits of merging
BBHs therefore depend on several of the varied evolutionary prescriptions
as well as on the initial binary parameter distributions.

\textbf{Formation spins:}
The natal-kick magnitude prescription has the largest \emph{all-BBH} total-effect
index for every spin response shown: 0.85, 0.73 and 0.75 for the
p10, median and p90 of $\chi_{\mathrm{eff}}$, and 0.87 for median
$\chi_{\mathrm{p}}$. In the merger ensemble, however, the leading
MEP differs between spin summaries. The lower $\chi_{\mathrm{eff}}$
percentile has its largest Shapley effect for natal-kick direction (0.27),
followed by common-envelope efficiency (0.22) and the initial binary
parameter distributions (0.21). The angular-momentum-loss parameter
$\beta$ has the largest allocation for the median and upper percentile
(0.29 and 0.30), with appreciable contributions from the kick magnitude
prescription and the initial distributions.
For median $\chi_{\mathrm{p}}$, the kick magnitude prescription remains
the leading MEP (0.41), followed by the initial distributions (0.30).
The different rankings show why
effective-spin percentiles and precession spin provide complementary
descriptions of MEP dependence; a ranking based on one spin summary
does not apply to the entire spin distribution.

Magnetic braking (\texttt{MB}) and the fallback threshold for full natal kicks
(\texttt{kickback}) have
total-effect indices below 0.01 for all ten \emph{all-BBH} responses.
Their merger Shapley effects are also small, reaching approximately
0.04 and 0.05, respectively, for median delay time. This is a statement
about the tested MEP values or prescriptions and selected summaries, rather than their
importance for every binary or evolutionary channel. More generally,
the heatmaps identify which MEPs contribute to variation across
models, while Fig.~\ref{fig:bbh-factor-responses} supplies the direction
of changes for its displayed median responses. The different rankings
in the two heatmaps describe their respective populations and variance
allocations; they do not by themselves isolate the effect of the
merger cut.

\section{Discussion}
\label{sec:discussion}

The model grid shows that the properties of Pop~III BBHs depend jointly
on the initial binary parameter distributions, the adopted evolutionary
prescriptions and the population being considered. Merger selection
changes more than the number of binaries: among the 992 models containing
at least ten mergers, it lowers median total mass in 792 and raises
median mass ratio in 940. These shifts indicate that the most massive
BBHs formed in a model need not be representative of its merging
population. The contrasting birth-eccentricity distributions further
show that the merger requirement can select different orbital regimes.
The mass distribution of merging BBHs therefore depends on both the
remnant masses produced by stellar evolution and which of these binaries
have orbits that allow them to merge within $t_{\mathrm H}$.

The correlation results extend this population dependence to relationships
between properties. In particular, the negative association between
aligned and precession spin becomes much less common after merger
selection, while the negative chirp-mass--mass-ratio tail is populated
by an entirely different set of models.

Considering several correlations together can identify more specific
combinations of MEP values and prescriptions within the tested grid. The joint requirements
$\rho_s(q,\chi_{\mathrm{eff}})<-0.3$ and
$\rho_s(\chi_{\mathrm{eff}},\chi_{\mathrm{p}})>0.3$, for example,
select 32 \emph{all-BBH} models and 20 different merger models, with
different natal-kick prescriptions in the two groups. Their association
with particular prescriptions motivates studying the joint
mass and spin distributions when comparing evolutionary models.
The chosen correlation cutoffs help identify patterns within the tested
grid. Although the selected models share particular MEP values or prescriptions, this
does not establish that an observed correlation pattern would uniquely
identify those MEP values or prescriptions, especially when other prescriptions and
observational uncertainties are considered. The physical links discussed
in Sect.~\ref{sec:results-correlations} identify possible roles for mass
transfer, remnant formation and recoil; the present analysis does not
determine which evolutionary sequences produce the observed signs.

The sensitivity analysis supplies a complementary view of parameter
dependence by measuring variation of population summaries across models.
The initial binary parameter distributions dominate the \emph{all-BBH}
mass-ratio, chirp-mass and birth-orbit responses, whereas the natal-kick
magnitude prescription dominates the delay-time and formation-spin
responses. In the merger ensemble, the initial distributions remain
central to the mass summaries, while common-envelope efficiency also
contributes substantially to mass ratios, birth orbits and delays.
The MEP with the largest contribution also differs between spin
properties and between the median and percentiles of their distributions:
kick direction has the largest allocation for the lower aligned-spin percentile,
angular-momentum loss during mass transfer for its median and upper
percentile, and kick magnitude for median precession spin.

The scope of these results is set by the discrete grid and the
adopted isolated-binary framework. Equal configuration weights describe
the explored models without assigning astrophysical probabilities to
them. The variation does not encompass uncertainties in assumptions
held fixed, including the initial BH spin of 0.05. 
Finite binary samples also limit the precision of population summaries and coefficients. The
at-least-ten-merger requirement used for the detailed comparisons does
not guarantee precision, and the sensitivity analysis includes all
1053 nonempty merger models, including the sparsest samples.

Predicting an observable population would introduce further uncertainties
from the Pop~III star-formation history and the fractions of stars in
single, binary and higher-order multiple systems, which remain poorly
constrained \citep{KlessenGlover2023}. Evolution in higher-order systems
also lies outside the isolated-binary framework. Keeping these uncertainties
separate allows the present study to identify how the adopted binary
prescriptions shape intrinsic properties. Testing whether the identified
correlation patterns help distinguish models observationally will require
star-formation and delay weighting, detector selection effects, measurement
uncertainties and contributions from other formation channels. Models
incorporating Pop~III star-formation histories and predictions for
detectable BBHs will be presented by Singh et al.\ (in preparation).

\section{Conclusions}
\label{sec:conclusions}
We have explored isolated Pop~III binary evolution with \textsc{StarTrack}
using 1152 configurations spanning eight MEPs, with $2\times10^6$
initial binaries per model. Comparing the \emph{all-BBH} population
with systems whose total delay does not exceed $t_{\mathrm H}$, we
find the following:
\begin{itemize}
  \item \textbf{Merger selection changes the mass distribution.}
    Among the 992 models containing at least ten mergers, selection
    lowers median total mass in 792 and raises median mass ratio in
    940. Model medians of total mass span $34.3$--$171.0\,M_{\odot}$
    across the 1152 \emph{all-BBH} models and
    $23.1$--$94.8\,M_{\odot}$ across these 992 merger models.

  \item \textbf{Merging populations span different orbital and spin properties.}
    Within the 992-model merger sample, median birth eccentricities
    range from $0.039$ to $0.983$, so merging BBHs do not share a
    single birth-eccentricity regime. Median aligned spins are positive,
    ranging from $0.018$ to $0.129$, while median precession spins
    range from zero to $0.094$. These are formation-spin values under
    the adopted spin prescription and initial BH spin of 0.05.

  \item \textbf{Correlations change when merging binaries are selected.}
    The negative association between aligned and precession spin is less
    common among \emph{merging-BBH} models than among \emph{all-BBH}
    models. The models with negative chirp-mass--mass-ratio
    coefficients below $-0.3$ are entirely different in the two
    populations. Comparisons within the same model configurations
    show that this change is not caused by excluding models with
    too few mergers.

  \item \textbf{Joint correlation patterns identify particular combinations of MEP values and prescriptions.}
    Requiring both $\rho_s(q,\chi_{\mathrm{eff}})<-0.3$ and
    $\rho_s(\chi_{\mathrm{eff}},\chi_{\mathrm{p}})>0.3$ selects
    32 \emph{all-BBH} models and 20 different merger models.
    This joint correlation pattern is associated with different natal-kick
    prescriptions in the \emph{all-BBH} and \emph{merging-BBH} populations.
    Such patterns provide candidates for
    distinguishing evolutionary models; they do not yet establish
    that observations can uniquely identify their MEP values or prescriptions.

  \item \textbf{Initial binary parameter distributions are central to mass and orbital variation.}
    They have the largest Sobol total effects for the mass-ratio,
    chirp-mass and birth-orbit statistics in the \emph{all-BBH} grid.
    In the nonempty merger ensemble, they receive the largest Shapley
    allocations for the chirp-mass and mass-ratio statistics, while
    common-envelope efficiency also contributes substantially to mass ratios, birth
    orbits and delays. These findings refer to the two ensembles and
    their respective variance measures.

  \item \textbf{Spin sensitivity depends on the property and percentile examined.}
    Natal-kick magnitude has the largest Sobol total effect for all
    formation-spin statistics studied in the \emph{all-BBH} grid.
    For merging BBHs, natal-kick direction has the largest Shapley
    allocation for the 10th percentile of aligned spin, angular-momentum
    loss during mass transfer for its median and 90th percentile,
    and natal-kick magnitude for median precession spin.
\end{itemize}
These conclusions describe intrinsic populations within the tested grid.
Assessing their implications for observed BBHs requires Pop~III
star-formation and merger-delay weighting, detector selection effects,
measurement uncertainties and contributions from other formation channels.
The data underlying this study, together with the complete set of
model-distribution plots, will be made publicly available on Zenodo.

\begin{acknowledgements}
This work was supported by the Universitat de les Illes Balears (UIB) with funds from the
Programa de Foment de la Recerca i la Innovació de la UIB 2024-2026 (supported by the
yearly plan of the Tourist Stay Tax ITS2023-086); the Spanish Agencia Estatal de
Investigación grants PID2022-138626NB-I00, RED2024-153978-E, RED2024-153735-E,
funded by MICIU/AEI/10.13039/501100011033 and the ERDF/EU; and the Comunitat
Autònoma de les Illes Balears through an “Ajut per a projectes de recerca científica i tecnològica” (PRD2025\_100) and through the Conselleria d’Educació i Universitats with funds from the European Regional Development Fund (SINCO2022/18146 - Plataforma
HiTech-IAC3-BIO). NS is financed by the Conselleria d’Educació i Universitats del Govern
de les Illes Balears and the European Social Fund Plus (ESF+) 2021-2027, through grant
POSTDOC2024\_55 and is also supported by Programa “Beques Santander - Investigació
Postdoctoral”, and by COST action SCALES CA24139, supported by COST (European
Cooperation in Science and Technology). TB acknowledges funding from the National
Science Centre, Poland (grant No. 2023/49/B/ST9/02777). NS acknowledges the computing support from the Astronomical Observatory, University of Warsaw, and the computer resources (Picasso Supercomputer and Magerit), technical expertise and assistance provided by Universidad Politécnica de Madrid and the SCBI (Supercomputing and Bioinformatics) center of the University of Málaga (AECT-2026-1-0022, AECT-2026-1-0011, and AECT-2026-1-0076).

The authors used OpenAI Codex (GPT-6) to assist with editing the manuscript
based on analyses, results and scientific interpretations provided by the
authors, developing and modifying plotting scripts, and checking consistency
between the manuscript and saved analysis outputs.
The authors retain full responsibility for the scientific interpretation
and final content of the manuscript.
\end{acknowledgements}

\bibliographystyle{aa}
\bibliography{references}

@article{AasiEtAl2015LIGO,
  author = {Aasi, J. and Abbott, B. P. and Abbott, R. and others},
  collaboration = {LIGO Scientific Collaboration},
  title = {{Advanced LIGO}},
  journal = {Classical and Quantum Gravity},
  year = {2015},
  volume = {32},
  pages = {074001},
  doi = {10.1088/0264-9381/32/7/074001},
  eprint = {1411.4547},
  archivePrefix = {arXiv}
}

@article{AcerneseEtAl2015Virgo,
  author = {Acernese, F. and Agathos, M. and Agatsuma, K. and others},
  collaboration = {Virgo Collaboration},
  title = {{Advanced Virgo: A Second-generation Interferometric Gravitational Wave Detector}},
  journal = {Classical and Quantum Gravity},
  year = {2015},
  volume = {32},
  pages = {024001},
  doi = {10.1088/0264-9381/32/2/024001},
  eprint = {1408.3978},
  archivePrefix = {arXiv}
}

@article{AkutsuEtAl2021KAGRA,
  author = {Akutsu, T. and Ando, M. and Arai, K. and others},
  collaboration = {KAGRA Collaboration},
  title = {{Overview of KAGRA: Detector Design and Construction History}},
  journal = {Progress of Theoretical and Experimental Physics},
  year = {2021},
  volume = {2021},
  pages = {05A101},
  doi = {10.1093/ptep/ptaa125},
  eprint = {2005.05574},
  archivePrefix = {arXiv}
}

@article{AbbottEtAl2019GWTC1Population,
  author = {Abbott, B. P. and Abbott, R. and Abbott, T. D. and others},
  collaboration = {LIGO Scientific Collaboration and Virgo Collaboration},
  title = {{Binary Black Hole Population Properties Inferred from the First and Second Observing Runs of Advanced LIGO and Advanced Virgo}},
  journal = {\apjl},
  year = {2019},
  volume = {882},
  pages = {L24},
  doi = {10.3847/2041-8213/ab3800},
  eprint = {1811.12940},
  archivePrefix = {arXiv}
}

@article{AbbottEtAl2020GW190521,
  author = {Abbott, R. and Abbott, T. D. and Abraham, S. and others},
  collaboration = {LIGO Scientific Collaboration and Virgo Collaboration},
  title = {{GW190521: A Binary Black Hole Merger with a Total Mass of $150\,M_{\odot}$}},
  journal = {Physical Review Letters},
  year = {2020},
  volume = {125},
  pages = {101102},
  doi = {10.1103/PhysRevLett.125.101102},
  eprint = {2009.01075},
  archivePrefix = {arXiv}
}

@article{AbbottEtAl2024GWTC21,
  author = {Abbott, R. and Abbott, T. D. and Acernese, F. and others},
  collaboration = {LIGO Scientific Collaboration and Virgo Collaboration},
  title = {{GWTC-2.1: Deep Extended Catalog of Compact Binary Coalescences Observed by LIGO and Virgo During the First Half of the Third Observing Run}},
  journal = {Physical Review D},
  year = {2024},
  volume = {109},
  pages = {022001},
  doi = {10.1103/PhysRevD.109.022001},
  eprint = {2108.01045},
  archivePrefix = {arXiv}
}

@article{AbbottEtAl2023GWTC3,
  author = {Abbott, R. and Abbott, T. D. and Acernese, F. and others},
  collaboration = {LIGO Scientific Collaboration and Virgo Collaboration and KAGRA Collaboration},
  title = {{GWTC-3: Compact Binary Coalescences Observed by LIGO and Virgo during the Second Part of the Third Observing Run}},
  journal = {Physical Review X},
  year = {2023},
  volume = {13},
  pages = {041039},
  doi = {10.1103/PhysRevX.13.041039},
  eprint = {2111.03606},
  archivePrefix = {arXiv}
}

@article{AbbottEtAl2023GWTC3Population,
  author = {Abbott, R. and Abbott, T. D. and Acernese, F. and others},
  collaboration = {LIGO Scientific Collaboration and Virgo Collaboration and KAGRA Collaboration},
  title = {{Population of Merging Compact Binaries Inferred Using Gravitational Waves through GWTC-3}},
  journal = {Physical Review X},
  year = {2023},
  volume = {13},
  pages = {011048},
  doi = {10.1103/PhysRevX.13.011048},
  eprint = {2111.03634},
  archivePrefix = {arXiv}
}

@article{AbacEtAl2025GW231123,
  author = {Abac, A. G. and Abouelfettouh, I. and Acernese, F. and others},
  collaboration = {LIGO Scientific Collaboration and Virgo Collaboration and KAGRA Collaboration},
  title = {{GW231123: A Binary Black Hole Merger with Total Mass $190$--$265\,M_{\odot}$}},
  journal = {\apjl},
  year = {2025},
  volume = {993},
  pages = {L25},
  doi = {10.3847/2041-8213/ae0c9c},
  eprint = {2507.08219},
  archivePrefix = {arXiv}
}

@article{BrommLarson2004,
  author = {Bromm, Volker and Larson, Richard B.},
  title = {{The First Stars}},
  journal = {\araa},
  year = {2004},
  volume = {42},
  pages = {79--118},
  doi = {10.1146/annurev.astro.42.053102.134034},
  eprint = {astro-ph/0311019},
  archivePrefix = {arXiv}
}

@article{KlessenGlover2023,
  author = {Klessen, Ralf S. and Glover, Simon C. O.},
  title = {{The First Stars: Formation, Properties, and Impact}},
  journal = {\araa},
  year = {2023},
  volume = {61},
  pages = {65--130},
  doi = {10.1146/annurev-astro-071221-053453},
  eprint = {2303.12500},
  archivePrefix = {arXiv}
}

@article{RyuEtAl2016,
  author = {Ryu, Taeho and Tanaka, Takamitsu L. and Perna, Rosalba},
  title = {{Formation, Disruption and Energy Output of Population III X-ray Binaries}},
  journal = {\mnras},
  year = {2016},
  volume = {456},
  pages = {223--238},
  doi = {10.1093/mnras/stv2629},
  eprint = {1509.05427},
  archivePrefix = {arXiv}
}

@article{BelczynskiEtAl2002,
  author = {Belczynski, Krzysztof and Kalogera, Vassiliki and Bulik, Tomasz},
  title = {{A Comprehensive Study of Binary Compact Objects as Gravitational Wave Sources: Evolutionary Channels, Rates, and Physical Properties}},
  journal = {\apj},
  year = {2002},
  volume = {572},
  pages = {407--431},
  doi = {10.1086/340304},
  eprint = {astro-ph/0111452},
  archivePrefix = {arXiv}
}

@article{BelczynskiEtAl2008,
  author = {Belczynski, Krzysztof and Kalogera, Vassiliki and Rasio, Frederic A. and Taam, Ronald E. and Zezas, Andreas and Bulik, Tomasz and Maccarone, Thomas J. and Ivanova, Natalia},
  title = {{Compact Object Modeling with the StarTrack Population Synthesis Code}},
  journal = {\apjs},
  year = {2008},
  volume = {174},
  pages = {223--260},
  doi = {10.1086/521026},
  eprint = {astro-ph/0511811},
  archivePrefix = {arXiv}
}

@article{BelczynskiEtAl2017,
  author = {Belczynski, Krzysztof and Ryu, Taeho and Perna, Rosalba and Berti, Emanuele and Tanaka, Takamitsu L. and Bulik, Tomasz},
  title = {{On the Likelihood of Detecting Gravitational Waves from Population III Compact Object Binaries}},
  journal = {\mnras},
  year = {2017},
  volume = {471},
  pages = {4702--4721},
  doi = {10.1093/mnras/stx1759}
}

@article{SantoliquidoEtAl2023,
  author = {Santoliquido, Filippo and Mapelli, Michela and Iorio, Giuliano and Costa, Guglielmo and Glover, Simon C. O. and Hartwig, Tilman and Klessen, Ralf S. and Merli, Lorenzo},
  title = {{Binary Black Hole Mergers from Population III Stars: Uncertainties from Star Formation and Binary Star Properties}},
  journal = {\mnras},
  year = {2023},
  volume = {524},
  pages = {307--324},
  doi = {10.1093/mnras/stad1860},
  eprint = {2303.15515},
  archivePrefix = {arXiv}
}

@article{TanikawaEtAl2021,
  author = {Tanikawa, Ataru and Susa, Hajime and Yoshida, Takashi and Trani, Alessandro A. and Kinugawa, Tomoya},
  title = {{Merger Rate Density of Population III Binary Black Holes Below, Above, and in the Pair-instability Mass Gap}},
  journal = {\apj},
  year = {2021},
  volume = {910},
  pages = {30},
  doi = {10.3847/1538-4357/abe40d},
  eprint = {2008.01890},
  archivePrefix = {arXiv}
}

@article{Tanikawa2024,
  author = {Tanikawa, Ataru},
  title = {{Contribution of Population III Stars to Merging Binary Black Holes}},
  journal = {Reviews of Modern Plasma Physics},
  year = {2024},
  volume = {8},
  pages = {13},
  doi = {10.1007/s41614-024-00153-8},
  eprint = {2403.04389},
  archivePrefix = {arXiv}
}

@article{TanikawaEtAl2026,
  author = {Tanikawa, Ataru and Liu, Shuai and Wu, WeiWei and Fujii, Michiko S. and Wang, Long},
  title = {{GW231123 Formation from Population III Stars: Isolated Binary Evolution}},
  journal = {\apj},
  year = {2026},
  volume = {1003},
  pages = {80},
  doi = {10.3847/1538-4357/ae62f5},
  eprint = {2508.01135},
  archivePrefix = {arXiv}
}

@article{IvanovaTaam2003,
  author = {Ivanova, Natalia and Taam, Ronald E.},
  title = {{Magnetic Braking Revisited}},
  journal = {\apj},
  year = {2003},
  volume = {599},
  pages = {516--521},
  doi = {10.1086/379192},
  eprint = {astro-ph/0308361},
  archivePrefix = {arXiv}
}

@article{FryerEtAl2012,
  author = {Fryer, Chris L. and Belczynski, Krzysztof and Wiktorowicz, Grzegorz and Dominik, Michal and Kalogera, Vicky and Holz, Daniel E.},
  title = {{Compact Remnant Mass Function: Dependence on the Explosion Mechanism and Metallicity}},
  journal = {\apj},
  year = {2012},
  volume = {749},
  pages = {91},
  doi = {10.1088/0004-637X/749/1/91},
  eprint = {1110.1726},
  archivePrefix = {arXiv}
}

@article{BelczynskiEtAl2016PI,
  author = {Belczynski, K. and Heger, A. and Gladysz, W. and Ruiter, A. J. and Woosley, S. and Wiktorowicz, G. and Chen, H.-Y. and Bulik, T. and O'Shaughnessy, R. and Holz, D. E. and Fryer, C. L. and Berti, E.},
  title = {{The Effect of Pair-instability Mass Loss on Black-hole Mergers}},
  journal = {\aap},
  year = {2016},
  volume = {594},
  pages = {A97},
  doi = {10.1051/0004-6361/201628980},
  eprint = {1607.03116},
  archivePrefix = {arXiv}
}

@article{HobbsEtAl2005,
  author = {Hobbs, G. and Lorimer, D. R. and Lyne, A. G. and Kramer, M.},
  title = {{A Statistical Study of 233 Pulsar Proper Motions}},
  journal = {\mnras},
  year = {2005},
  volume = {360},
  pages = {974--992},
  doi = {10.1111/j.1365-2966.2005.09087.x}
}

@article{BrayEldridge2016,
  author = {Bray, J. C. and Eldridge, J. J.},
  title = {{Neutron Star Kicks and Their Relationship to Supernovae Ejecta Mass}},
  journal = {\mnras},
  year = {2016},
  volume = {461},
  pages = {3747--3759},
  doi = {10.1093/mnras/stw1275}
}

@article{Peters1964,
  author = {Peters, P. C.},
  title = {Gravitational Radiation and the Motion of Two Point Masses},
  journal = {Physical Review},
  year = {1964},
  volume = {136},
  pages = {B1224--B1232},
  doi = {10.1103/PhysRev.136.B1224}
}

@article{SchmidtEtAl2015,
  author = {Schmidt, Patricia and Ohme, Frank and Hannam, Mark},
  title = {Towards models of gravitational waveforms from generic binaries: {II}. Modelling precession effects with a single effective precession parameter},
  journal = {Physical Review D},
  year = {2015},
  volume = {91},
  pages = {024043},
  doi = {10.1103/PhysRevD.91.024043},
  eprint = {1408.1810},
  archivePrefix = {arXiv}
}

@article{GerosaEtAl2021,
  author = {Gerosa, Davide and Mould, Matthew and Gangardt, Daria and Schmidt, Patricia and Pratten, Geraint and Thomas, Lucy M.},
  title = {A generalized precession parameter {$\chi_{\mathrm{p}}$} to interpret gravitational-wave data},
  journal = {Physical Review D},
  year = {2021},
  volume = {103},
  pages = {064067},
  doi = {10.1103/PhysRevD.103.064067},
  eprint = {2011.11948},
  archivePrefix = {arXiv}
}

@article{Sobol2001,
  author = {Sobol', I. M.},
  title = {Global sensitivity indices for nonlinear mathematical models and their {Monte Carlo} estimates},
  journal = {Mathematics and Computers in Simulation},
  year = {2001},
  volume = {55},
  pages = {271--280},
  doi = {10.1016/S0378-4754(00)00270-6}
}

@article{Owen2013,
  author = {Owen, Art B.},
  title = {Variance Components and Generalized {Sobol'} Indices},
  journal = {SIAM/ASA Journal on Uncertainty Quantification},
  year = {2013},
  volume = {1},
  pages = {19--41},
  doi = {10.1137/120876782}
}

@article{SongEtAl2016,
  author = {Song, Eunhye and Nelson, Barry L. and Staum, Jeremy},
  title = {{Shapley} Effects for Global Sensitivity Analysis: Theory and Computation},
  journal = {SIAM/ASA Journal on Uncertainty Quantification},
  year = {2016},
  volume = {4},
  pages = {1060--1083},
  doi = {10.1137/15M1048070}
}

@article{OwenPrieur2017,
  author = {Owen, Art B. and Prieur, Cl{\'e}mentine},
  title = {On {Shapley} Value for Measuring Importance of Dependent Inputs},
  journal = {SIAM/ASA Journal on Uncertainty Quantification},
  year = {2017},
  volume = {5},
  pages = {986--1002},
  doi = {10.1137/16M1097717}
}

@article{CostaEtAl2023,
  author = {Costa, Guglielmo and Mapelli, Michela and Iorio, Giuliano and Santoliquido, Filippo and Escobar, Gast{\'{o}}n J and Klessen, Ralf S and Bressan, Alessandro},
  title = {{Massive binary black holes from Population II and III stars}},
  year = {2023},
  journal = {\mnras},
  volume = {525},
  number = {2},
  pages = {2891--2906},
  doi = {10.1093/mnras/stad2443},
  eprint = {2303.15511},
  archivePrefix = {arXiv},
  url = {https://arxiv.org/abs/2303.15511}
}

@article{TanikawaEtAl2022,
  author = {Tanikawa, Ataru and Yoshida, Takashi and Kinugawa, Tomoya and Trani, Alessandro A. and Hosokawa, Takashi and Susa, Hajime and Omukai, Kazuyuki},
  title = {{Merger Rate Density of Binary Black Holes through Isolated Population I, II, III and Extremely Metal-poor Binary Star Evolution}},
  journal = {\apj},
  year = {2022},
  volume = {926},
  pages = {83},
  doi = {10.3847/1538-4357/ac4247},
  eprint = {2110.10846},
  archivePrefix = {arXiv}
}

@article{HijikawaEtAl2022,
  author = {Hijikawa, Kotaro and Kinugawa, Tomoya and Tanikawa, Ataru and Yoshida, Takashi and Umeda, Hideyuki},
  title = {{On the population III binary black hole mergers with intermediate mass black holes: dependence on common envelope parameter}},
  journal = {arXiv e-prints},
  year = {2022},
  pages = {arXiv:2211.07496},
  doi = {10.48550/arXiv.2211.07496},
  eprint = {2211.07496},
  archivePrefix = {arXiv}
}

\begin{appendix}
\section{Variance-based sensitivity analysis}
\label{app:variance-analysis}

\subsection{Responses and model weights}
\label{app:variance-ensemble}
For a specified binary
population and property, each configuration provides one response $y_m$,
such as its median chirp mass $\mathcal{M}$ or a percentile of
$\chi_{\mathrm{eff}}$. The variance analysed
here is therefore the variance of these summaries across models, rather
than the variance of individual binary properties within a model. 
We combine the two generation sets into a catalogue of 1152 configurations;
the ensemble used for each analysis is specified below.

Let $\boldsymbol{x}_m$ contain the eight varied MEPs of model $m$:
$\alpha$, $\beta$, \texttt{MB}, \texttt{REMNANT}, \texttt{SS},
\texttt{kick}, \texttt{kickback} and \texttt{polar} (see Table~\ref{tab:input-parameters}).
Let $\mathcal{I}=\{1,\ldots,d\}$, with $d=8$, index these MEPs. On an
ensemble $\mathcal{E}$ of $N$ configurations, we assign probability $1/N$
to each pair $(\boldsymbol{x}_m,y_m)$. The ensemble mean
$\mu_{\mathcal{E}}$ and the between-model variance $V_{\mathcal{E}}$
of the response are then
\begin{equation}
 \mu_{\mathcal{E}}=\frac{1}{N}\sum_{m\in\mathcal{E}}y_m,
 \qquad
 V_{\mathcal{E}}=\frac{1}{N}\sum_{m\in\mathcal{E}}
                (y_m-\mu_{\mathcal{E}})^2.
 \label{eq:ensemble-variance}
\end{equation}
The denominator is $N$ because these are moments of the specified discrete
distribution. A model producing more binaries receives no additional
weight. These weights describe the sampled grid; they are not
astrophysical occurrence probabilities. We analyse each response
separately on its original scale, conditional on the realised simulation
summaries.

\subsection{Full-grid decomposition and Sobol indices}
\label{app:sobol}
For the \emph{all-BBH} summaries, including systems that do not merge within
the adopted Hubble-time threshold $t_{\mathrm H}$,
all $N=1152$ configurations have defined responses. Their MEP values or prescriptions
form a complete Cartesian grid
$\mathcal{X}=\mathcal{X}_1\times\cdots\times\mathcal{X}_d$.
Here $\mathcal{X}_i$ is the set of tested values or prescriptions for
MEP $i$, and $n_i=|\mathcal{X}_i|$ is the number of these MEP values or prescriptions.
For example, $\alpha\in\{0.5,1,1.5\}$ has three values.
Equal model weights then imply
\begin{equation}
 p(\boldsymbol{x})=\prod_{i=1}^{d}\frac{1}{n_i},
 \qquad N=\prod_{i=1}^{d}n_i=1152.
 \label{eq:grid-measure}
\end{equation}
This product distribution permits an orthogonal Sobol variance
decomposition, retaining both contributions from individual MEPs and
interactions
\citep{Sobol2001,Owen2013}.

If a model configuration is selected according to the probabilities in
Eq.~(\ref{eq:grid-measure}), its eight MEP values or prescriptions define a random
vector $\boldsymbol{X}$. A particular combination of these MEP values or prescriptions is
denoted by the lowercase vector $\boldsymbol{x}$. If $f$ assigns a response to each
combination, then $f(\boldsymbol{x}_m)=y_m$ for model $m$. 
The response associated with the randomly selected
configuration is then the random variable $Y=f(\boldsymbol{X})$.
For a subset $u\subseteq\mathcal{I}$ of MEP indices, let $\bar u$ denote its
complement, containing all indices not in $u$. Thus,
$\boldsymbol{x}_u$ contains the selected MEPs and
$\boldsymbol{x}_{\bar u}$ contains the remaining MEPs.
Holding the selected MEPs fixed and averaging over all tested
combinations of the remaining MEPs gives the conditional grid average
\begin{equation}
 g_u(\boldsymbol{x}_u)
 =\mathrm{E}[Y\mid\boldsymbol{X}_u=\boldsymbol{x}_u]
 =\frac{1}{\prod_{j\in\bar u}n_j}
   \sum_{\boldsymbol{x}_{\bar u}} f(\boldsymbol{x}_u,\boldsymbol{x}_{\bar u}).
 \label{eq:conditional-grid-mean}
\end{equation}
Here $\mathrm{E}[\cdot]$ denotes expectation (the weighted mean) under
the grid distribution, and the vertical bar specifies the MEP values or prescriptions
held fixed.

Let $\varnothing$ denote the empty set of MEP indices. When
$u=\varnothing$, no MEPs are held fixed, so $g_{\varnothing}$ is the
mean response over all 1152 configurations, $\mu_{\mathcal{X}}$.
We express each model's response as this overall mean plus contributions
from individual MEPs and their interactions. This overall mean provides
a common baseline from which the response of each configuration can differ.
We denote this baseline by
$f_{\varnothing}=g_{\varnothing}=\mu$, with $\mu=\mu_{\mathcal{X}}$.
For a nonempty subset $u$, the component associated with precisely its
MEPs is
\begin{equation}
 f_u(\boldsymbol{x}_u)=g_u(\boldsymbol{x}_u)
                -\sum_{v\subsetneq u}f_v(\boldsymbol{x}_v).
 \label{eq:anova-recursion}
\end{equation}
For example, $f_i=g_i-\mu$ and
$f_{ij}=g_{ij}-g_i-g_j+\mu$. Thus an interaction measures the part of
the joint response that remains after its lower-order components have
been removed. Evaluating Eq.~(\ref{eq:anova-recursion}) through all
$2^8-1=255$ nonempty subsets gives the exact grid identity
\begin{equation}
 f(\boldsymbol{x})=\mu+
        \sum_{\varnothing\ne u\subseteq\mathcal{I}}f_u(\boldsymbol{x}_u).
 \label{eq:anova-expansion}
\end{equation}
No interpolation or assumed polynomial dependence between numerical MEP values
is required.

By construction, the contributions $f_u$ for nonempty subsets have zero
mean and are mutually orthogonal: the product of any two distinct
contributions averages to zero over the grid. Consequently, the total
variance is the sum of their individual variances,
\begin{equation}
 V=\sum_{\varnothing\ne u\subseteq\mathcal{I}}V_u,
 \qquad
 V_u=\frac{1}{N}\sum_{\boldsymbol{x}\in\mathcal{X}}
                         f_u(\boldsymbol{x}_u)^2,
 \label{eq:variance-components}
\end{equation}
where $V=V_{\mathcal{X}}$. For $V>0$, the first-order and total-effect
indices are
\begin{align}
 S_i&=\frac{V_{\{i\}}}{V}
     =\frac{\operatorname{Var}(\mathrm{E}[Y\mid X_i])}{V},
       \label{eq:sobol-first}\\
 S_{\mathrm{T},i}&=\frac{1}{V}\sum_{u\ni i}V_u
     =\frac{\mathrm{E}[\operatorname{Var}(Y\mid\boldsymbol{X}_{\bar i})]}{V}.
       \label{eq:sobol-total}
\end{align}
Here $\bar i$ denotes the set of all MEP indices except $i$, so
$\boldsymbol{X}_{\bar i}$ contains all other MEPs. In the second
expression in Eq.~(\ref{eq:sobol-total}), we hold these other MEPs fixed
and calculate the response variance as MEP $i$ varies over its tested
values or prescriptions. Averaging this variance over all combinations of the other MEPs
gives the total variance contribution involving $i$, including its
interactions.
The \emph{all-BBH} heatmap in Fig.~\ref{fig:bbh-variance-heatmaps}(a)
displays $S_{\mathrm{T},i}$, so interactions enter
the total of each participating MEP. In particular,
\begin{equation}
 \sum_{i=1}^{d}S_{\mathrm{T},i}
       =\sum_{\varnothing\ne u\subseteq\mathcal{I}}|u|\frac{V_u}{V}.
 \label{eq:total-index-sum}
\end{equation}
Individual indices lie between zero and one, but their sum exceeds one
whenever a nonzero interaction component is present. We do not
renormalise these rows.

\subsection{Shapley effects for the merger-selected ensemble}
\label{app:shapley}
Merger-selected summaries use binaries with total formation-to-merger
delay $\leq t_{\mathrm H}$. Models with no BBHs merging within this
threshold have undefined distribution summaries and are therefore excluded.
Our catalogue contains $N_{\mathrm{M}}=1053$ models with at least one
qualifying BBH merger; the remaining 99 models have none and are excluded.
The models with nonempty merger samples include 61 with only one to nine mergers.
Equal weight is assigned to each model in this ensemble.
Removing configurations generally breaks the product structure in
Eq.~(\ref{eq:grid-measure}), inducing dependence among MEPs.
We therefore use variance-based Shapley effects on the joint distribution of MEPs in this ensemble
\citep{SongEtAl2016,OwenPrieur2017}.

Let $\mathcal{E}_{\mathrm{M}}$ denote this ensemble and let
$\mu_{\mathrm{M}}$ and $V_{\mathrm{M}}$ be its moments from
Eq.~(\ref{eq:ensemble-variance}), using its merger-selected responses.
For each subset $u$, partition the models in this ensemble into groups
$\mathcal{G}_u$ that have identical values or prescriptions for the MEPs in $\boldsymbol{X}_u$.
Writing $n_g$ for the number of models in group $g$ and
$\bar y_g=n_g^{-1}\sum_{m\in g}y_m$, the explained-variance fraction is
\begin{equation}
 C(u)=\frac{\operatorname{Var}_{\mathrm{M}}
                  (\mathrm{E}_{\mathrm{M}}[Y\mid\boldsymbol{X}_u])}
                {V_{\mathrm{M}}}
     =\frac{\sum_{g\in\mathcal{G}_u}n_g(\bar y_g-\mu_{\mathrm{M}})^2}
                 {N_{\mathrm{M}}V_{\mathrm{M}}}.
 \label{eq:shapley-value-function}
\end{equation}
Each group is weighted by its fraction of the models in this ensemble,
$n_g/N_{\mathrm{M}}$, ensuring that every model contributes equally.
For $V_{\mathrm{M}}>0$, $C(\varnothing)=0$ and
$C(\mathcal{I})=1$, since each complete MEP vector identifies one
tabulated response.

For each MEP $i$, we calculate the increase in the explained-variance
fraction when it is added to a subset $u$ of the other MEPs.
Writing $u\cup\{i\}$ for the subset with $i$ added, the Shapley effect
$\phi_i$ is the weighted average of these increases:
\begin{equation}
 \phi_i=\sum_{u\subseteq\mathcal{I}\setminus\{i\}}
 \frac{|u|!\,(d-|u|-1)!}{d!}
 \left[C(u\cup\{i\})-C(u)\right].
 \label{eq:shapley-effects}
\end{equation}
Here $|u|$ is the number of MEPs in $u$. The standard Shapley weights
give a normalised allocation with nonnegative effects that sum to unity
\citep{SongEtAl2016,OwenPrieur2017}:
\begin{equation}
 \phi_i\geq0,\qquad \sum_{i=1}^{d}\phi_i=1.
 \label{eq:shapley-normalisation}
\end{equation}
We evaluate all $2^8=256$ subset values directly from the saved responses
and use the subset sum in Eq.~(\ref{eq:shapley-effects}).

\subsection{Interpretation and limitations}
\label{app:variance-interpretation}
For independent MEPs, $C(u)=\sum_{\varnothing\ne v\subseteq u}V_v/V$. An interaction
component $V_v$ enters a permutation increment when the last of its
$|v|$ participating MEPs is added; each is equally likely to be last.
Consequently,
\begin{equation}
 \phi_i=\sum_{v\ni i}\frac{V_v}{|v|V}
 \qquad\hbox{(independent MEPs)}.
 \label{eq:shapley-independent}
\end{equation}
Shapley effects thus divide each interaction among its participants,
whereas total-effect Sobol indices include it in full for each one
\citep{OwenPrieur2017}. On the merger-selected ensemble the Shapley
allocation also reflects information shared through MEP dependence.
The two heatmaps differ in population, model support and allocation
rule; their differences cannot be attributed solely to the merger cut.
\end{appendix}

\end{document}